\documentclass[%
 reprint,
superscriptaddress,
showpacs,preprintnumbers,
 amsmath,amssymb,
 aps,
 prl,
]{revtex4-1}
\usepackage{slashed}
\usepackage{graphicx}
\usepackage{dcolumn}
\usepackage{bm}

\begin{document}

\title{The Zeeman, Spin-Orbit, and Quantum Spin Hall Interactions in Anisotropic and Low-Dimensional Conductors }

\author{Aiying Zhao}
\affiliation{Department of Physics, University of Science and Technology Beijing,  Beijing 100083, China}
\affiliation{Department of Physics, University of Central Florida, Orlando, FL 32816-2385, USA}
\author{Qiang Gu}
\email{qgu@ustb.edu.cn, corresponding author}
\affiliation{Department of Physics, University of Science and Technology Beijing,  Beijing 100083, China}
\author{Timothy J. Haugan}
\affiliation{U. S. Air Force Research Laboratory, Wright-Patterson Air Force Base, Ohio 45433-7251, USA}
\author{Richard A. Klemm}
\email{richard.klemm@ucf.edu, corresponding author}
\affiliation{Department of Physics, University of Central Florida, Orlando, FL 32816-2385, USA}
\affiliation{U. S. Air Force Research Laboratory, Wright-Patterson Air Force Base, Ohio 45433-7251, USA}
\date{\today}

\begin{abstract} When an electron is free or in the ground state of an atom, its $g$-factor is 2, as first shown by Dirac.  But when an electron or hole is in a conduction band of a crystal, it can be very different from 2, depending upon the crystalline anisotropy and the direction of the applied magnetic induction ${\bm B}$.  In fact, it can even be 0!  To demonstrate this quantitatively, the Dirac equation is extended for a relativistic electron or hole in an orthorhombically-anisotropic conduction band with effective masses $m_j$ for $j=1,2,3$ with geometric mean $m_g=(m_1m_2m_3)^{1/3}$. Its covariance is established with
general proper and improper Lorentz transformations. The appropriate Foldy-Wouthuysen transformations are extended to evaluate the non-relativistic Hamiltonian
to $O({\rm m}c^2)^{-4}$, where ${\rm m}c^2$ is the particle's Einstein rest energy.  The results can have extremely important consequences for magnetic measurements   of many classes of clean anisotropic
semiconductors, metals, and superconductors.  For  ${\bm B}||\hat{\bm e}_{\mu}$, the Zeeman $g_{\mu}$ factor is  $2{\rm m}\sqrt{m_{\mu}}/m_g^{3/2} + O({\rm m}c^2)^{-2}$.  While propagating in a  two-dimensional (2D) conduction band with $m_3\gg m_1,m_2$,  $g_{||}<<2$, consistent with recent measurements of the temperature $T$ dependence of the parallel upper critical induction $B_{c2,||}(T)$ in superconducting monolayer NbSe$_2$ and in twisted bilayer graphene. While a particle is in its conduction band of an atomically thin
one-dimensional metallic chain along $\hat{\bm e}_{\mu}$,  $g<<2$   for all ${\bm B}={\bm\nabla}\times{\bm A}$ directions and vanishingly small for ${\bm B}||\hat{\bm e}_{\mu}$.    The quantum spin Hall Hamiltonian for 2D metals with $m_1=m_2=m_{||}$ is $K[{\bm E}\times({\bm p}-q{\bm A})]_{\perp}\sigma_{\perp}+O({\rm m}c^2)^{-4}$, where ${\bm E}$ and ${\bm p}-q{\bm A}$ are the planar  electric field and gauge-invariant momentum, $q=\mp|e|$ is the particle's charge,  $\sigma_{\perp}$ is the Pauli matrix normal to the layer, $K=\pm\mu_B/(2m_{||}c^2)$, and $\mu_B$ is the Bohr magneton.

\end{abstract}

 \vskip0pt
\maketitle


\section{Introduction}
What is the electron $g$-factor? When an electron is in an atom, its response to an applied magnetic induction ${\bm B}$ leads to the linear Zeeman energy $H_Z=\frac{|e|}{2{\rm m}}({\bm L}+2{\bm S})\cdot{\bm B}$, plus corrections of order ${\bm B}^2$, where ${\bm L}$ and ${\bm S}$ are the electron's orbital and spin angular momenta, ${\rm m}$ and $-e$ are its rest mass and charge,  the $g$-factor is the Land{\'e} $g$-factor $g_J$, which depends upon the atomic quantum numbers $\ell$, $s$, and $j$ \cite{Feynman,Griffiths}, and  the factor of 2 in the numerator of $H_Z$ is a relativistic result first derived by Dirac \cite{Dirac}. If there were no spin, the electron $g$-factor would be 1 \cite{Feynman}. When an electron is in the ground $^2$S$_{1/2}$ state of a hydrogen atom, $\ell=0$, and the Land{\'e} $g$-factor is the ``pure spin'' value of 2 \cite{Feynman}. If it were in a $^{2}$P$_{1/2}$ state of a hydrogen atom, $g_J=2/3$ due to spin-orbit coupling. But what is the $g$-factor when the electron is not confined to an atom, but propagates in a conduction band of a crystal?  What is it if the crystal is a monolayer of graphene or NbSe$_2$, in metallic and/or superconducting twisted bilayer graphene, or a metallic single-walled carbon nanotube?  What is it for a hole in its crystalline conduction band?  These are fundamental questions that have been completely ignored in essentially all papers written on semiconductors, metals, and superconductors. Here we present for the first time a rigorous answer to all of these questions.  The results imply that drastic changes in the interpretations of many existing experiments are necessary, and many theoretical treatments of lower dimensional conductors need to be redone.  In particular, the Zeeman, spin-orbit, and quantum spin Hall interactions of an electron or hole depend strongly upon the anisotropy and/or dimensionality of the conduction band.

In semiconductors such as Si and Ge, and in metallic Bi, the lowest energy conduction bands have ellipsoidal symmetry, with $\epsilon_0({\bm p})=\sum_{i=1}^3p_i^2/(2m_i)$
about some point in the first Brillouin zone  \cite{Cohen,CohenBlount,Mahan}.  The normal state of the high-temperature superconductor YBa$_2$CuO$_{7-\delta}$ (YBCO) is metallic
in all three orthorhombic directions \cite{Friedmann,Klemmbook}, so that it is reasonable to assume its $\epsilon_0({\bm p})$ has that form, with $m_3\gg m_1, m_2$.  Although standard adiabatic Knight shift measurements on the planar
 $^{63}$Cu nuclei of that material showed the standard, temperature-dependent behavior through the superconducting transition temperature $T_c$ when the strong time-independent magnetic induction ${\bm B}_0\perp\hat{\bm c}$ (perpendicular to the direction of mass $m_3$) and the oscillatory in time component ${\bm B}_1(t)$ was perpendicular to that direction, measurements with ${\bm B}_0||\hat{\bm c}$ and ${\bm B}_1(t)\perp{\bm B}_0$ showed anomalous, temperature-independent behavior through the superconducting state \cite{Barrett}. Those measurements were inconsistent with the standard model  that neglects any anisotropy features of the Fermi surface and of the Zeeman
interaction \cite{Yosida}.   Further questions about  Knight shift
measurements were raised by the apparent incompatibility of pulsed, non-adiabatic $^{17}$O Knight shift measurements with ${\bm B}_0\perp\hat{\bm c}$ with scanning tunneling measurements of the superconducting gap in the
highly layered Sr$_2$RuO$_4$ \cite{Ishida,Suderow}, although very recent adiabatic $^{17}$O Knight shift measurements on that material with  ${\bm B}_0$ in the same direction  were consistent with a singlet spin state in that material \cite{Pustogow,Ishida2}, as were the Knight shift measurements in the same directions on YBCO.  The Knight shift dimensionality issue still
remains unresolved in the quasi-one-dimensional organic superconductor (TMTSF)$_2$PF$_6$, where TMTSF is tetramethyl-tetraselenofulvalene\cite{Chaikin}, for which strong
dimensionality issues could arise for both the strong constant ${\bm B}_0$ and the weaker perpendicular ${\bm B}_1(t)$.

Now excellent quality lower-dimensional conductors can be prepared in the clean limit \cite{Cao}, so it is important to reassess the  effects of the crystal anisotropy upon the conduction particle's Zeeman energy.
Here we present a theory of a relativistic electron or hole in an orthorhombically anisotropic conduction band, and demonstrate its invariance  under the most general proper and improper Lorentz
transformations.  We also employ a modified form of the Klemm-Clem transformations to transform it to isotropic form \cite{Klemmbook,KlemmClem}, in which its invariance under each of the improper transformations of charge conjugation, parity, and time reversal (CPT) is easily established.
Before those transformations,   the appropriate Foldy-Wouythuysen transformations of the anisotropic Hamiltonian are greatly extended to evaluate
its non-relativistic limit  to order $({\rm m}c^2)^{-4}$ \cite{FW}, where ${\rm m}c^2$ is the particle's
Einstein rest energy, in order to investigate the dimensionality of the Zeeman, spin-orbit, and quantum spin Hall interactions with great precision.

We found that the Zeeman interaction only exists for
${\bm B}$ normal to the conducting plane for an electron or hole while travelling in a two-dimensional conduction band, and there is no Zeeman, spin-orbit, and quantum spin Hall interactions in one dimension for any ${\bm B}$ direction, {\it provided that}
the probed particle is moving within the one-dimensional conduction band.  These results lead to the absence of Pauli limiting of the upper critical field $B_{c2,||}(T)$
parallel to clean ultrathin superconductors, as  observed recently \cite{Cao,Xi,Lu,Fatemi,Sajadi,Agosta,Matsuda} and the Pauli-limiting effects upon $B_{c2,\perp}(T)$
normal to the film are strongly related to the conduction particle's effective mass within the conducting plane \cite{KLB,FF,LO}.  The results also have profound implications for
interpretations of Knight shift measurements on highly anisotropic superconductors such as YBCO, Sr$_2$RuO$_4$, and (TMTSF)$_2$PF$_6$,
\cite{Ishida,Pustogow,Ishida2,Chaikin,HallKlemm,magnetochemistry}. They  also have important consequences of magnetic field studies of the metallic states of other quasi-one-dimensional conductors such as metallic single-walled carbon nanotubes \cite{Zeinab,Dresselhaus}, tetrathiafulvalene tetracyanoquinodimethane (TTF-TCNQ), K$_2$Pt(CN)$_4$Br$_{0.3}\cdot$3H$_2$O, (SN)$_x$, NbSe$_3$, ZrTe$_5$, o-TaS$_3$, K$_{0.3}$MoO$_3$ and related compounds \cite{Keller,1Dconductors,Hokkaido}. The results also have important consequences for very thin samples of quasi-two-dimensional conductors and semiconductors, such as many twisted bilayer graphene, monolayer FeSe, transition metal dichalcogenides, including semimetallic TiS$_2$, metallic and superconducting NbSe$_2$, TaS$_2$, WTe$_2$, MoS$_2$,  and their intercalates \cite{Cao,He,Xi,Lu,Fatemi,Sajadi,Klemmbook,KlemmPristine}, and layered heavy fermion conductors, \cite{Agosta,Matsuda},{\it etc.}  The interaction of the spin of an electron or hole with  ${\bm B}_0+{\bm B}_1(t)$ is greatly
reduced during the particle's low-dimensional conduction, compared to that while confined to an atom.

 The strong reduction of the Zeeman interaction $g$-factor for ${\bm B}$ parallel to a clean two-dimensional superconductor can quantitatively account for the extreme violation of the Pauli limit in monolayer NbSe$_2$ and gated MoS$_2$ \cite{Xi,Lu}, without  invoking Ising or models involving strong spin-orbit coupling\cite{Xi,Lu,Wang}.  Moreover, we provide a simple explanation for the temperature $T$ dependence of the upper critical induction $B_{c2,||}(T)$ parallel to the twisted bilayers of superconducting graphene \cite{Cao}: the $g$ factor for that field direction is extremely small.  This cannot be explained by Ising or spin-orbit coupling models.

\section{The Model}

When an electron or hole is bound to an atomic nucleus, a zero-dimensional (0D) case, as sketched in Fig. 1(a), its localized motion about that nucleus is fully three-dimensional (3D) on an atomic scale, as for the electron in the hydrogen atom, for which its spin experiences a full 3D Zeeman effect with ${\bm B}$, as derived  from the isotropic Dirac equation for an electron in the presence of a radial electrostatic potential \cite{Dirac,BjorkenDrell}.  However, from the moment that electron or hole leaves that atom and travels in a one-dimensional (1D) conduction band, as sketched in Fig. 1(b), its delocalization drastically changes its interaction with ${\bm B}$, so that there is no Zeeman interaction during that delocalized 1D motion.

\begin{figure}
\center{\includegraphics[width=0.45\textwidth]{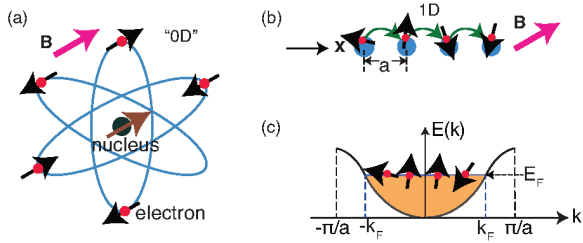}
\caption{ Atomic (``0D'') spin components with a full $H_{3D}^Z$ and 1D electronic or hole conduction with $H_{1D}^Z=0$. (a) Sketch of an atom of effective point size (``0D''), in which the nuclear components and the electrons move in localized atomic 3D environments. (b) An electron or hole moving on a 1D chain with $H_{1D}^Z=0$. (c) A tight-binding conduction band model for 1D motion, with the delocalized continuous energy levels $E(k)$ filled up to the Fermi energy $E_F$.  The inverse of the effective electron-like or hole-like effective mass is given by the curvature at $E_F$ of such a conduction band.}}
\end{figure}

However, for a very large number $N\sim10^8$ of atomic sites in the conducting chain,  the energy states $E(k)$ in the tight-binding model of 1D motion form a  continuous band, as sketched in Fig. 1(c), and the electrons or holes at the Fermi energy $E_F$ move with velocities $\pm v_F$, the magnitude of which can be on the order of  $10^{-2}c$, where $c$ is the speed of light in vacuum, so its motion is much more like that of a relativistic particle in a 1D world with regard to its interaction with ${\bm B}$ than like that of a localized atomic electron.  The inverse of its effective mass $m_1$ is determined from the curvature of $E(k)$ evaluated at $E_F$.  If $E_F$ is less that one-half of the bandwidth, the curvature at $E_F$ is positive, and the motion is electron-like.  If $E_F$ is greater than 1/2 the bandwidth, then the curvature at $E_F$ is negative, and the conduction is hole-like. In both cases, its motion can be treated in the non-relativistic limit of the one-dimensional (1D) Schr{\"o}dinger equation based upon the  Hamiltonian

\begin{eqnarray}
H_{1D}&=&\sqrt{{\rm m}c^2\Pi_{1}^2/m_{1}+{\rm m}^2c^4}+q\Phi(x,t),
\end{eqnarray}
where $\Pi_{1}=p_{x}-qA_{x}$, ${\rm m}$ is the particle's rest mass, $q$ is its charge, which is $-|e|$ for an electron or $+|e|$ for a hole, $p_{x}=-i\hbar\frac{\partial}{\partial x}$ is the particle's momentum, and $A_{x}(x,t)$ is the magnetic vector potential in its 1D world,
respectively, where $\hbar$ is Planck's constant divided by 2$\pi$.  In Coulomb gauge, $\frac{\partial A_x}{\partial x}=0$, so $A_x$ is only a function of the time $t$.  As Dirac showed for an electron in a three-dimensional (3D) world \cite{Dirac}, one can linearize this Hamiltonian using two Pauli matrices.
Even without employing the Pauli matrices, there cannot be a vector product or a curl  in a 1D world, so it is therefore easy to see that the magnetic induction ${\bm
B}={\bm \nabla}\times{\bm A}$ must vanish, and there cannot  be any Zeeman energy while the particle is travelling in its 1D conduction band.

 In two dimensions (2D) with
effective mass $m_{||}=m_1=m_{2}$, the particle's effective Hamiltonian may be written as
\begin{eqnarray}
H_{2D}&=&\sqrt{{\rm m}c^2(\Pi_1^2+\Pi_2^2)/m_{||}+{\rm m}^2c^4}+q\Phi(x,y,t),
\end{eqnarray}
where $\Pi_1=-i\hbar\frac{\partial}{\partial x}-qA_x(x,y,t)$, $\Pi_2=-i\hbar\frac{\partial}{\partial y}-qA_y(x,y,t)$, and in Coulomb gauge in the 2D world,
$\frac{\partial A_x}{\partial x}+\frac{\partial A_y}{\partial y}=0$, and ${\bm B}$ can have  a non-vanishing component normal to the conducting
plane, as sketched in Figs. 2(a) and 2(b), but both of its components within the 2D conducting plane vanish, as sketched in Figs. 2(c) and 2(d). In Fig. 3(a), a planar cross-section of the spin-split electron energy dispersions shown in Fig. 2(b) is shown, leading to the 2D electron  Fermi surfaces sketched in Fig. 3(b). On the other hand, for an ``inverted'' hole-like band, the spin-split Fermi surfaces are sketched in Fig. 3(c), and in materials with both electron and hole bands, such as many transition metal dichalcogenides \cite{Klemmbook,KlemmPristine}, both spin-split electron and hole Fermi surfaces can exist, provided that ${\bm B}$ is normal to the 2D conduction plane, a tetragonal version of which is sketched in Fig. 3(d).

Although these arguments regarding the dimensionality of the Zeeman interaction are very
simple, a correct theoretical analysis of the appropriate Dirac equations for these lower dimensional conducting bands does indeed lead to the same conclusions:  There are
no Zeeman, spin-orbit, or quantum spin Hall interactions of a particle while moving in a 1D conduction band, and the ${\bm B}$ field with which a particle interacts while traveling in a 2D conduction band is precisely normal to the conducting plane.

\begin{figure}
\center{\includegraphics[width=0.45\textwidth]{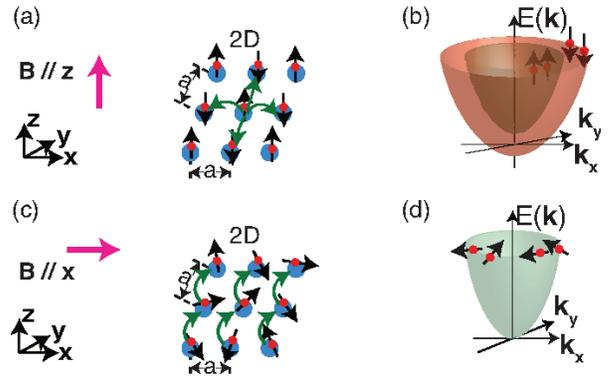}
\caption{ 2D motion and its highly anisotropic $H^Z_{2D}$. (a) Sketch of a 2D ionic lattice in the $xy$ plane with ${\bm B}||{\bm
z}$.  The electron  spins experience a full $H^{Z}_{2D,\perp}$.  (b) Sketches of $E({\bm k})$ for both spins parallel and antiparallel to ${\bm B}||{\bm z}$. (c) Sketch of the same 2D ionic lattice with ${\bm B}||{\bm x}$. The electron spins have $H^{Z}_{2D,||}=0$. (d)  Sketch of the single $E({\bm k})$ for both
spin states with ${\bm B}||{\bm x}$.
}}
\end{figure}

\begin{figure}
\includegraphics[width=0.45\textwidth]{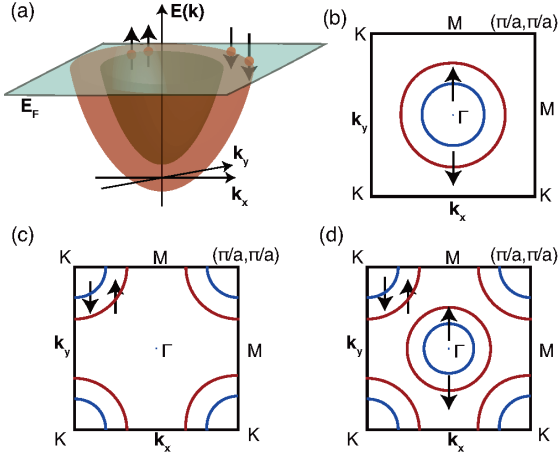}
\caption{Typical 2D particle and hole bands. (a) Sketch of a plane at finite $E$ that intersects the two electron Fermi surfaces split by the Zeeman effect with ${\bm B}||\hat{\bm z}$. (b) Sketch of the 2D electron Fermi surfaces for a low band filling with ${\bm B}||\hat{\bm z}$. (c) Sketch of the two hole Fermi surfaces obtained with higher band filling with ${\bm B}||\hat{\bm z}$. (d) Sketch of a 2D system with both electron and hole Fermi surfaces split by the Zeeman effect with ${\bm B}||\hat{\bm z}$. }
\end{figure}

More generally, a relativistic electron or hole in an anisotropic environment of orthorhombic symmetry satisfies the Schr{\"o}dinger equation based upon the modified
Hamiltonian $\tilde{H}={\tilde T}+\varepsilon$, where  $\varepsilon({\bm r},t)=q\Phi({\bm r},t)$ and
 \begin{eqnarray}
 {\tilde T}&=&\sqrt{{\rm m}c^2\sum_{\mu=1}^3\Pi_{\mu}^2/m_{\mu}+{\rm m}^2c^4},
 \end{eqnarray}
where $\Pi_{\mu}=p_{\mu}-qA_{\mu}({\bm r},t)$, and $m_{\mu}$, $p_{\mu}$ and $A_{\mu}$ are the components of the particle's effective mass, its momentum, and the magnetic
vector potential, respectively, and we assume the time $t$ dependencies of the scalar and vector potentials $\Phi({\bm r},t)$ and ${\bm A}({\bm r},t)$ are slow with
respect to differences in the particle's energies divided by $\hbar$, so that they can be treated adiabatically.

Dirac first showed that one can linearize the isotropic version of this Hamiltonian by the use of matrices based upon the Pauli matrices \cite{Dirac}, and generalizing it
to
orthorhombic anisotropy, we have
\begin{eqnarray}
\tilde{H}\psi&=&\left[\tilde{\cal O}+\beta {\rm m}c^2+\varepsilon\right]\psi=i\hbar\frac{\partial\psi}{\partial t},\\
\tilde{\cal O}&=&c\tilde{\bm\alpha}\cdot{\bm \Pi},
\end{eqnarray}
where ${\bm\Pi}={\bm p}-q{\bm A}$, ${\bm p}\rightarrow-i\hbar{\bm\nabla}$, and
\begin{eqnarray}
\tilde{\alpha}_{\mu}&=&\left[\begin{array}{cc}0&\tilde{\sigma}_{\mu}\\
\tilde{\sigma}_{\mu}&0\end{array}\right],\hskip5pt\tilde{\sigma}_{\mu}=\frac{\sigma_{\mu}}{\sqrt{m_{\mu}/{\rm
m}}},\hskip5pt\beta=\left[\begin{array}{cc}1&0\\0&-1\end{array}\right],
\end{eqnarray}
for $\mu=1,2,3$, the $\sigma_{\mu}$ are the Pauli matrices, and both the $\tilde{\alpha}_{\mu}$ and $\beta$ are rank-4 matrices, where 1 represents the rank-2 identity
matrix
\cite{Dirac,BjorkenDrell}.

These four traceless matrices satisfy $\left\{\tilde{\alpha}_{\mu},\tilde{\alpha}_{\nu}\right\}=2\delta_{\mu\nu}{\rm m}/m_{\mu}$, and $\left\{\tilde{\alpha}_{\mu},\beta\right\}=0$.

From Eq. (4), the $\mu_{\rm th}$ component of the probability  current  is   $j_{\mu}=\psi^{\dagger}\tilde{\alpha}_{\mu}\psi$, and since $\rho=\psi^{\dag}\psi$, the
continuity
equation
$\frac{\partial\rho}{\partial t}+\frac{\partial}{\partial x_{\mu}}j_{\mu}=\frac{\partial\rho}{\partial t}+{\rm div}{\bm j}=0$,
 is still satisfied with effective mass anisotropy.

\subsection{\label{sec:level2}Proof of covariance}

\begin{figure}
\center{\includegraphics[width=0.45\textwidth]{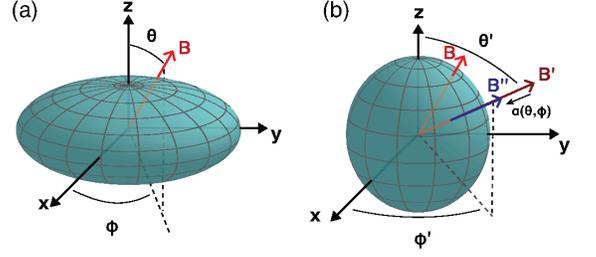}
\caption{Sketch of the effects of the special case of a modified Klemm-Clem transformations for a constant ${\bm B}$. (a) The untransformed ellipsoidal Fermi surface for an orthorhombically anisotropic conduction band, with a constant magnetic induction ${\bm B}$ in an arbitrary direction ${\bm B}=B(\sin\theta\cos\phi,\sin\theta\sin\phi,\cos\theta)$. (b) After the anisotropic scale transformation, the Fermi surface is transformed to a sphere, and the transformed magnetic induction ${\bm B}'$ differs both in magnitude and direction from ${\bm B}$.  Then, the isotropic scale transformation of ${\bm B}'$ to ${\bm B}''||{\bm B}'$ leaves its direction invariant, but changes it magnitude $|{\bm B}'|=B$ to that of the untransformed ${\bm B}$, as described in the Appendix.} }
\end{figure}

Details of the most general proper Lorentz transformation (a rotation about all three orthogonal axes and a boost to a general special relativistic reference frame moving
at a constant velocity in an arbitrary direction) are posted in Appendix.  However, a much simpler proof is given here.
We use a version of the Klemm-Clem transformations that were used to transform  an orthorhombically anisotropic Ginzburg-Landau model of an anisotropic superconductor
into isotropic form \cite{KlemmClem}.  To do so, we first make the anisotropic scale transformation of the spatial parts of the contravariant form of the anisotropic
Dirac equation, Eq. (4),
\begin{eqnarray}
\frac{\partial}{\partial x_{\mu}}&=&\sqrt{m_{\mu}/{\rm m}}\frac{\partial}{\partial x_{\mu}'},\\
A_{\mu}&=&\sqrt{m_{\mu}/{\rm m}}A_{\mu}',
\end{eqnarray}
which transforms Eq. (1) to
\begin{eqnarray}
i\hbar\frac{\partial\psi}{\partial t}&=&\left[c{\bm\alpha}\cdot{\bm\Pi}'+\beta {\rm m}c^2+q\Phi\right]\psi=H\psi,
\end{eqnarray}
where
\begin{eqnarray}
\alpha_{\mu}&=&\left[\begin{array}{cc}0&\sigma_{\mu}\\ \sigma_{\mu}&0\end{array}\right],
\end{eqnarray}
which is precisely the same form as the isotropic Dirac equation.

It is easy to show  that this transformation preserves the Maxwell equation of no monopoles, ${\bm\nabla}'\cdot{\bm B}'=0$, provided that
\begin{eqnarray}
B_{\mu}&=&\overline{C}\sqrt{{\rm m}/m_{\mu}}B_{\mu}',
\end{eqnarray}
for $\mu=1,2,3$,
where $\overline{C}$ can be any  constant.  Then it is easy to show that this form preserves the required relation
\begin{eqnarray}
{\bm B}'&=&{\bm\nabla}'\times{\bm A}',
\end{eqnarray}
provided that
\begin{eqnarray}
\overline{C}&=&(m_g/{\rm m})^{3/2},
\end{eqnarray}
where $m_g=(m_1m_2m_3)^{1/3}$ is the geometric mean effective mass.
We note that ${\bm B}'$ is no longer parallel to ${\bm B}$, but these transformations are fully general for arbitrary ${\bm A}$ and ${\bm B}$, which can depend upon position in accordance with Maxwell's equations.

 In the special case of ${\bm B}$ in a fixed direction, ${\bm B}'$ can be made parallel to ${\bm B}$ by a proper rotation \cite{Klemmbook,KlemmClem}.  The magnitude $|{\bm B}'|$ can also be made equal to
$|{\bm B}|$ by an isotropic scale transformation \cite{Klemmbook,KlemmClem}, which preserves the isotropy of the transformed Dirac equation, as described in the Appendix, and as pictured in Fig. 4.

Improper Lorentz transformations such as reflections are represented by rank-4 matrices $b$ satisfying ${\rm det}(b)=-1$.  As discussed in the Appendix, a general rotation $\tilde{R}$ with arbitrary effective mass anisotropy is found to be a rank-4 matrix $\tilde{R}=e^{-{\bm\omega}\cdot\tilde{\bm S}}$, the determinant of which is +1, appropriate for a proper Lorentz transformation, as is the determinant of a general boost with full effective mass anisotropy $\tilde{B}=e^{-\tilde{\bm\zeta}\cdot\tilde{\bm K}}$.  Thus, based upon the general theorem of the determinant $|AB|=|A||B|=|BA|$ of the product of two matrices of rank-$n$ \cite{Nering}, so that the order of rotation and reflection is irrelevant.

Similar to isotropic case, Eq. (4) is invariant under  charge conjugation, parity, and time-reversal (CPT) transformations and we have showed it in Appendix.
\subsection{\label{sec:level3} Expansion about the non-relativistic Limit}

In order to explore the low energy properties of the anisotropic Dirac equation, we extended the Foldy-Wouthuysen transformations \cite{FW,BjorkenDrell} to eliminate the odd terms
in the anisotropic operator $\tilde{{\cal O}}=c\tilde{\bm\alpha}\cdot{\bm\Pi}$
obtained in the power series in $({\rm m}c^2)^{-1}$  to the non-relativistic limit of $\tilde{H}$ in Eq. (4). Since the initial release of these results expanded to order $({\rm m}c^2)^{-2}$ shocked many workers \cite{Zhao}, we decided to carry out the expansion to order $({\rm m}c^2)^{-4}$, in order to strengthen our arguments. Setting $\tilde{\Gamma}=[\tilde{\cal{O}},\epsilon]+i \hbar
\frac{\partial{\tilde{ \cal{O}}}}{\partial t}$ $= i\hbar cq\tilde{\bm{\alpha}} \cdot {\bm{E}({\bm r},t)}$, ${ \varepsilon}=q\Phi({\bm r},t)$, ${\bm E}({\bm r},t)=-{\bm\nabla}\Phi({\bm r},t)-\frac{\partial{\bm A}({\bm r},t)}{\partial t}$,  $\mu_{B}=q\hbar/(2{\rm m})$ is the Bohr magneton for holes and
minus the Bohr magneton for electrons. We note that the commutator $[\beta,\varepsilon]\equiv\beta\varepsilon-\varepsilon\beta=0$ and the anticommutator $\{\tilde{\cal O},\beta\}\equiv\tilde{\cal O}\beta+\beta\tilde{\cal O}=0$. To order $({\rm m}c^2)^{-4}$,
\begin{eqnarray}
\tilde{H}_{3D}&=& \beta \Biggl( {\rm m}c^2+ \frac{1}{2{\rm{m}}c^2}\tilde{{\cal O}}^2  -\frac{1}{8{\rm{m}}^3 c^6}\Bigl(\tilde{{\cal O}}^4
+\tilde{\Gamma}^2\Bigr)\nonumber\\
& &+\frac{i\hbar}{32{\rm m}^4c^6}\frac{\partial}{\partial t}\tilde{\Gamma}^2 \Biggr)-\frac{1}{8\rm{m}^2 c^4}[ \tilde{{\cal O}}, \tilde{\Gamma}]\\
&&+\frac{1}{384{\rm{m}}^4 c^8}\Bigl([\tilde{{\cal O}},[\tilde{{\cal O}},[\tilde{{\cal O}}, \tilde{\Gamma}]]]+12[\tilde{\Gamma},[\tilde{\Gamma},\varepsilon]]\Bigr)
 + \varepsilon.\nonumber
\end{eqnarray}
 We note that the fourth order term containing $[\tilde{\Gamma},\varepsilon]$ vanishes,  since
$\Phi({\bm r},t)$ commutes with each component of ${\bm E}({\bm r},t)$.   Although $\beta$ is a rank-4 matrix, it has only two diagonal elements, so that
the Hamiltonian for an electron or a hole in an anisotropic conduction band is obtained respectively \cite{BjorkenDrell,FW} with $\beta=1,-1$ and $q=-|e|,+|e|$.

We calculated $\tilde{H}_{3D}$ using the above procedure with the effective masses, and by first making the Klemm-Clem transformations in Eqs. (7), (8), and (11)-(13).  We note that  $\tilde{\cal O}=\tilde{\bm\alpha}\cdot{\bm\Pi}={\bm\alpha}\cdot{\bm\Pi}'$, and that $\tilde{\Gamma}=i\hbar cq\tilde{\bm\alpha}\cdot {\bm E}({\bm r},t)=i\hbar cq{\bm\alpha}\cdot {\bm E}'({\bm r}',t)$, where
\begin{eqnarray}
E_{\mu}({\bm r},t)&=&\sqrt{m_{\mu}/{\rm m}}E_{\mu}'({\bm r}',t),
\end{eqnarray}
where
\begin{eqnarray}
x_{\mu}&=&\sqrt{{\rm m}/m^{\mu}}x_{\mu}'.
\end{eqnarray}
We then may write
\begin{eqnarray}
H'_{3D}&=&\beta\Biggl({\rm m}c^2+\frac{{\bm \Pi'}^2}{2{\rm m}}-\mu_{B}{\bm \sigma}\cdot{\bm B}'+\frac{\mu_B^2}{2{\rm m}c^4}{{\bm E}'}^2 \nonumber \\
&&-\frac{1}{2{\rm m}c^2}\Bigl(\frac{{\bm \Pi'}^2}{2{\rm m}}-\mu_{B}{\bm \sigma}\cdot{\bm B}'\Bigr)^2-\frac{i\hbar\mu_B^2}{8{\rm m}^2c^6}\frac{\partial}{\partial t}({\bm E}')^2\Biggr)\nonumber\\
&&-\frac{\mu_{B}}{4{\rm m}c^2}\Bigl(\hbar{\bm \nabla}'\cdot{\bm E}'+(2{\bm E}'\times{\bm\Pi}'+i\hbar{\bm\nabla}'\times{\bm E}')\cdot{\bm\sigma}\Bigr)\nonumber\\
& &+\frac{\mu_B X'_{3D}}{192{\rm m}^3c^4}+q\Phi({\bm r}',t)+O({\rm m}c^2)^{-5},\label{H3DNR}
\end{eqnarray}
where $\mu_B=q\hbar/(2{\rm m})$ is the Bohr magneton for holes and minus the Bohr magneton for electrons, and $X_{3D}'$ is given in the Appendix.
  The spin-dependent  terms  in $H_{3D}'$ are
\begin{eqnarray}
&&H_{3D}^{Z'}=-\beta\mu_{B}\Bigl( {\bm \sigma}\cdot{\bm B}'-\frac{\{ {\bm \Pi'}^2,{\bm \sigma}\cdot{\bm B}'  \}}{4{\rm m^2}c^2}\Bigr)
         +\frac{\mu_B \sum\limits_{i=1}^{6}X_{3D,(i)}'}{192{\rm m}^3c^4},\label{H3DZ}\nonumber\\
         &&\\
&&H_{3D}^{SO'}=-\frac{\beta\mu_{B}}{4{\rm m}c^2}i\hbar{\bm\nabla}'\times{\bm E}'\cdot{\bm\sigma}+\frac{\mu_B\sum_{i=7}^{9}X'_{3D,(i)}}{192{\rm m}^3c^4},\label{H3DSO}\nonumber\\
& &\\
&&H_{3D}^{QSH'}=-\frac{\beta\mu_{B}}{2{\rm m}c^2}({\bm E}'\times{\bm\Pi}')\cdot {\bm \sigma}+\frac{\mu_B \sum\limits_{i=10}^{14}X'_{3D,(i)}}{192{\rm m}^3c^4},\label{H3DQSH}\\
&&H_{3D}^{Z,QSH'}=\frac{\mu_B X'_{3D,(15)}}{192{\rm m}^3c^4},\label{H3Dmixed}
\end{eqnarray}
where explicit forms for the $X'_{3D,(i)}$  are given in the Appendix.

It is then elementary to perform the inverse of the anisotropic scale transformations of ${\bm E}', {\bm \nabla}', {\bm A}'$, ${\bm\Pi}'$ and ${\bm M}'$.  Doing so leads to the result listed at the end of the Appendix that was obtained without first making the Klemm-Clem transformations in Eqs. (7), (8), (11)-(13), (15), and  (16).

We note that the Zeeman energy in a lower-dimensional conduction band is highly anisotropic, with the $g$ factor $g_{\mu}$ for ${\bm B}||\hat{\bm e}_{\mu}$ given by
\begin{eqnarray}
g_{\mu}&=&2\frac{{\rm m}\sqrt{m_{\mu}}}{m_g^{3/2}}(1-\delta_{g,\mu}),\nonumber\\
\delta_{g,\mu}&=&\frac{1}{4{\rm m}c^2}\sum_{\nu}\frac{1}{m_{\nu}}\frac{\left\{\Pi_{\nu}^2,B_{\mu}\right\}}{B_{\mu}}
                 -\frac{\beta\sum\limits_{i=6}^{11}X'_{3D,(i),\mu}}{192{\rm m}^3c^4 B'_{\mu}  }\nonumber\\
& &+O({\rm m}c^2)^{-4},\label{gnu}
\end{eqnarray}
for both electrons and holes,
where $\delta_{g,\mu}$ contains terms that operate upon the particle's wave function, and the $X'_{3D,(i),\mu}/B_{\mu}'$ must be transformed back to the laboratory frame, which can involve sums over the components of ${\bm E}, {\bm B}, {\bm\nabla}$, and ${\bm\Pi}$. Since the two lowest order contributions to $g_{\mu}$  in $({\rm m}c^2)^{-1}$ are both proportional to $\beta$, but the third order correction is not, we multiplied it by $\beta$ to compensate.  Through second  order, $g_{\mu}$ has the same sign for both electrons and holes, but in third order, they have opposite signs.

However, many  workers treated $g_{\mu}$ as having the isotropic value 2, and added that Zeeman energy by hand for an anisotropic system. Foldy and Wouthuysen included the
$E_{\mu}\Pi_{\nu}$ term to order $({\rm m}c^2)^{-2}$, but omitted the $i\hbar(\partial E_{\mu}/\partial x_{\nu})$ term \cite{FW}. Subsequently, Bjorken and Drell included both terms
\cite{BjorkenDrell}, but omitted the $A_{\nu}$ part of $\Pi_{\nu}$ in terms of second and third order in $1/({\rm m}c^2)$.  We emphasize that ${\bm A}$ and $\Phi$ are the important quantum mechanical potentials, as ${\bm B}$
can vanish in regions where ${\bm A}\ne0$, as noted in the text. The quantum spin Hall Hamiltonian for an anisotropic metal is given by Eq.(\ref{H3DQSH}). Note that $\tilde{H}_{QSH}$ explicitly contains $A_{\nu}$ in $\Pi_{\nu}$, a term that has consistently been dropped in many papers on the quantum spin Hall effect. \cite{QiZhang}

\section{Discussion}

Here we first describe the most important effects of motion dimensionality.  An electron or hole in an isotropic 3D conduction band with $m_1=m_2=m_3=m_g$ satisfies the
Schr{\"o}dinger equation $H'_{3D}\psi=i\hbar(\partial\psi/\partial t)$,  given Eq. (17) with ${\bm E}'\rightarrow \sqrt{{\rm m}/m_g}{\bm E}$, ${\bm\Pi}'\rightarrow\sqrt{{\rm m}/m_g}{\bm \Pi}$, and ${\bm B}'\rightarrow({\rm m}/m_g){\bm B}$.

For an electron or hole in a 2D metal with  $m_3\rightarrow\infty$, the anisotropic 2D Hamiltonian $H'_{2D}$ with $m_1\ne m_2$ to order $({\rm m}c^2)^{-4}$ is given in the Appendix.  For the isotropic 2D case $m_1=m_2=m_{||}$,   it reduces to
\begin{eqnarray}
H_{2D}&=&\beta\Biggl({\rm m}c^2+\frac{{\bm \Pi}^2_{||}}{2m_{||}}-\mu_{B||}\sigma_{\perp}B_{\perp}+\frac{\mu_{B}^2}{2m_{||}c^4}{\bm E}_{||}^2 \nonumber \\
&&-\frac{1}{2{\rm m}c^2}\Bigl(\frac{\bm \Pi^{2}_{||}}{2m_{||}}-\mu_{B||}\sigma_{\perp}B_{\perp}\Bigr)^2-\frac{i\hbar\mu_B^2}{8{\rm m}m_{||}c^6}(\frac{\partial}{\partial t}{\bm E}_{||}^2)\Biggr)\nonumber\\
&&-\frac{\mu_{B}}{4m_{||} c^2}\biggl(\Bigl(2{\bm E}\times{\bm\Pi}+i\hbar{\bm\nabla}\times{\bm
E}\Bigr)_{\perp}\sigma_{\perp}\nonumber\\
& & +\hbar({\bm \nabla}\cdot{\bm E})_{||}\biggr)\nonumber\\
& &+\frac{\mu_{B||}X_{2D}}{192{\rm m}^2m_{||} c^4} +q\Phi({\bm r}_{||},t)+O({\rm m}c^2)^{-5},
\end{eqnarray}
where$\mu_{B||}=q\hbar/(2m_{||})$ is the planar Bohr magneton, and $X_{2D}$ is given in the Appendix.
We note that the Zeeman, spin-orbit,  quantum spin Hall, and mixed terms may be written as
\begin{eqnarray}
&&H_{2D}^Z=-\beta\mu_{B||}( \sigma_{\perp}B_{\perp}-\frac{\{ {\bm \Pi}^2_{||},\sigma_{\perp}B_{\perp}  \}}{4{\rm mm_{||}}c^2})
         +\frac{\mu_{B||} \sum\limits_{i=1}^{3}X_{2D,(i)}}{192{\rm m}^2m_{||}c^4},\label{H2DZ}\nonumber\\
         & &\\
&&H_{2D}^{SO}=-\frac{\beta\mu_{B||}}{4{\rm m}c^2}i\hbar({\bm\nabla}\times{\bm E})_{\perp}\sigma_{\perp}+\frac{\hbar q \sum_{i=4}^5X_{2D,(i)}}{384{\rm m}^2m_{||}^2c^4},\label{H2DSO}\nonumber\\
& &\\
&&H_{2D}^{QSH}=-\frac{\beta\mu_{B||}}{2{\rm m}c^2}({\bm E}\times{\bm\Pi})_{\perp} {\bm \sigma}_{\perp}+\frac{\mu_B \sum\limits_{i=6}^{8}X_{2D,(i)}}{192{\rm m}m_{||}^2c^4},\label{H2DQSH}\\
&&H_{2D}^{Z,QSH}=\frac{\mu_{B||}X_{2D,{(9)}}}{192{\rm m}^2m_{||}c^4},
\end{eqnarray}
where the $X_{2D,(i)}$ are given in the Appendix.

The leading contribution to the 2D Zeeman interaction, $-\beta\mu_{B||}\sigma_{\perp}B'_{\perp}$,  is equivalent for electrons and holes,  and vanishes for ${\bm B}$ parallel to the infinitesimally thin 2D
metallic film.  We note again that in twisted-bilayer graphene and in monolayer NbSe$_2$, $B_{c2,||}(T)$ is consistent with  $g_{||}<<2$ \cite{Cao,Xi}.  This may also be the case in a large number of other clean 2D superconductors, including monolayer FeSe \cite{He}.

For an electron or hole in a one-dimensional conduction band with $m_1=m_{x}$, $m_2=m_3\rightarrow\infty$,
\begin{eqnarray}
H^{NR}_{1D}&=&\beta\Biggl({\rm m}c^2+\frac{\Pi^2_{x}}{2m_{x}}-\frac{\Pi_{x}^4}{8{\rm m}m_{x}^2c^2}+\frac{\mu_B^2}{2m_{x}c^4}E_{x}^2 \nonumber\\
&&\frac{i\hbar\mu_B^2}{{\rm m}m_x c^6}\frac{\partial}{\partial t} E_{x}^{2}\Biggr)-\frac{\hbar\mu_{B}}{4m_{x}c^2}\frac{\partial E_{x}}{\partial x}+q\Phi(x, t)\nonumber\\
& &-\frac{\hbar^3\mu_B }{192{\rm m}m_x^2 c^4}\frac{\partial^3 E_x}{\partial x^3} +O({\rm m}c^2)^{-5}.
\end{eqnarray}
None of the spin-orbit, quantum spin Hall, or Zeeman interactions exist in 1D, which we checked to order $({\rm
m}c^2)^{-4}$ using the orthorhombically-anisotropic form of the Foldy-Wouthuysen transformations.  Of course, in a quasi-one-dimensional metal, $g_{\mu}$ is given by Eq. (22).

\begin{figure}
\includegraphics[width=0.45\textwidth]{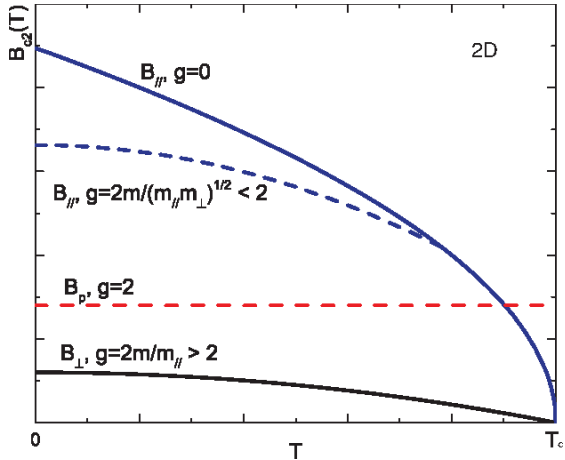}
\caption{Upper critical induction of a 2D superconductor.
 Sketches of the Tinkham ${\bm B}_{c2,||}(T)$ with $g=0$ (solid blue)\cite{Klemmbook,Tinkham}, an example of a weakly Pauli-limited ${\bm B}_{c2,||}(T)$ (short-dashed blue) with $g=2{\rm m}/(m_{||}m_{\perp})^{1/2}<2$, an example of a strongly Pauli-limited ${\bm B}_{c2,\perp}(T)$ with $g=2{\rm m}/m_{||}>2$ (solid black), and the conventional Pauli limit $B_{P}$ with $g=2$ (red dashed).}
\end{figure}

In models \cite{KLB,FF,LO}, the Zeeman interaction was assumed to be that of a free electron moving isotropically in three spatial dimensions (3D). On a macroscopic scale,
the size of an atom is a ''zero-dimensional'' (''0D'') point, as sketched in Fig. 1(a). Microscopically, however,  its nucleus  moves slowly inside a 3D electronic shell,
and as for the Dirac equation of a free electron, the 3D relativistic motion of each of its neutrons and protons leads to it having an overall  spin $I$ and a nuclear
Zeeman energy that can be probed by a time $t$-dependent external  ${\bm B}(t)$ in nuclear magnetic resonance (NMR) and in Knight shift measurements when in
a metal \cite{HallKlemm,magnetochemistry}.  The orbital electrons bound to that nucleus  also move in a nearly isotropic 3D environment,  and have a much  larger Zeeman
interaction with ${\bm B}(t)$, modified only by the $V({\bm r})=-e\Phi({\bm r})$  of nearby atoms.

However, when an atomic electron is excited into a crystalline conduction band, it leaves that atomic site and moves with wave vector ${\bm k}$ across the crystal. Its
motion depends upon the crystal structure, and can be highly anisotropic. In an isotropic, 3D metal,   $E({\bm k})=\hbar^2{\bm k}^2/(2{\rm m})$ for free electrons.  These
states are filled at $T=0$ up to the Fermi energy $E_F$ and $H^Z_{3D}=-\mu_B{\bm \sigma}\cdot{\bm B}$.  However, in   Si and Ge \cite{Mahan}, the lowest energy conduction
bands can be expressed as $E({\bm k})=\hbar^2\sum_{i=1}^3(k_i-k_{i0})^2/(2m_i)$ about some minimal point ${\bm k}_0$, and the $m_i$ can differ significantly from ${\rm m}$.

In a purely one-dimensional (1D) metal, the conduction electrons move rapidly along the chain of atomic sites, as sketched  in Fig. 1(b), usually with a tight-binding 1D
band $E(k)$ as sketched in Fig. 1(c),  and $H^{Z}_{1D}=0$.   When an electron is in a quasi-1D superconductor such as tetramethyl-tetraselenafulvalene
hexafluorophosphate, (TMTSF)$_2$PF$_6$ \cite{Chaikin}, $E({\bm k})$ is highly anisotropic, the transport normal to the conducting chains is  by weak hopping, so the effective
masses in those directions greatly exceed ${\rm m}$.

Similarly, in 2D metals, such as monolayer or gated NbSe$_2$, MoS$_2$, WTe$_2$ \cite{Xi,Lu,Fatemi,Sajadi}, and twisted bilayer graphene \cite{Cao}, the effective mass
normal to the conducting plane is effectively infinite.  As sketched in Fig. 2, the direction of ${\bm B}$ is very important.  When ${\bm H}$ is  normal to that plane, as
in Fig. 2(a), the spins of the conduction electrons eventually align either parallel or anti-parallel to ${\bm B}$, giving rise to a Zeeman interaction that can differ
from that of a free electron only by the effective mass $m_{||}$. There are two energy dispersions $E({\bm k})$ for up and down spin conduction electrons, as sketched in
Fig. 2(b).

However, when ${\bm B}$ lies within the 2D conduction plane, as sketched in Fig. 2(c), the Zeeman interactions vanish, so their spin states are effectively random, and
there is only one conduction band, as sketched in Fig. 2(d).

In Fig. 5, sketches of the generic behavior expected for the upper critical induction ${\bm B}_{c2}(T)$ for ${\bm B}=\mu_0{\bm H}$ applied parallel and perpendicular
to a 2D film.  The red dashed horizontal line is the effective Pauli limiting induction ${B}_P$, which is proportional to the effective mass $m_{||}$ within the
conducting plane, and can therefore be either larger or smaller than the result (1.86 $T_c$ T/K) for an isotropic  superconductor.  However, ${\bm B}_{c2,||}(T)$
generically follows the Tinkham thin film formula ${\bm B}_{c2,||}(T)=\mu_0\sqrt{3}\Phi_0/[\pi s\xi_{||}(T)]$ \cite{Klemmbook}, where $s$ is the film thickness,
$\Phi_0=h/(2e)$ is the superconducting flux quantum and $\xi_{||}(T)$ is the Ginzburg-Landau coherence length parallel to the film.  There is no Pauli limiting for this
${\bm B}$ direction, consistent with many  experiments \cite{Xi,Lu,Fatemi,Sajadi,Cao,Klemmbook}.

In an infinitessimally thin conducting layer, the Zeeman Hamiltonian to $O({\rm m}c^2)^{-4}$ is Eq.(\ref{H2DZ}).

The Knight shift is the relative change in the NMR frequency for a nuclear species when it is in a metal (or superconductor) from when it is in an insulator or vacuum.
In both cases, the nuclear spin of an atom interacts with that of  one of its orbital electrons via the hyperfine interaction.  But when that atom is in a metal, the
orbital electron can sometimes be excited into the conduction band, travelling throughout the crystal, and then returning to the same nuclear site, producing the leading
order contribution to the Knight shift \cite{HallKlemm,magnetochemistry}.  The dimensionality of the motion of the electron in the conduction band is therefore crucial in
interpreting Knight shift measurements of anisotropic materials, as first noticed in the anisotropic three-dimensional superconductor, YBa$_2$Cu$_3$O$_{7-\delta}$
\cite{Barrett}.

In Knight shift $K(T)$ measurements with  ${\bm B}$ applied parallel to the layers of  Sr$_2$RuO$_4$, $H^{Z}_{2D}$ should be vanishingly small, so one expects little
change in $K(T)$ at and below $T_c$, due to Eq. (3), as observed \cite{Ishida}. Similarly, Eq. (6) implies that $K(T)$ on the quasi-one-dimensional superconductor
(TMTSF)$_2$PF$_6$ should be nearly constant, as observed \cite{Chaikin}.    A recent $K(T)$ measurement on Sr$_2$RuO$_4$ under uniaxial planar pressure did show a substantial
$K(T)$ variation below $T_c$ \cite{Pustogow},  in agreement with scanning tunneling microscopy results of a nodeless superconducting gap \cite{Suderow}.

\section{Summary}

The Dirac equation  is extended to treat a relativistic charge with charge q in an orthorhombically anisotropic conduction band. The norm for this model with metric
$\tilde{g}$ is invariant under the most general proper Lorentz transformation $\tilde{A}$, the matrix representation of which exhibits  O(3,1) group symmetry, and this
anisotropic Dirac equation is demonstrated to be covariant, precisely as for the isotropic Dirac equation. This model applies to large classes of anisotropic
semiconductors, metals, and superconductors. Although overlooked by many workers, the  $\bm{A}$ in ${\bm \Pi}$ plays an important role in the quantum spin Hall Hamiltonian, which is distinctly different from that of
spin-orbital coupling in a topological insulator, and a proposed experiment to test this result will be published elsewhere.

This model has profound consequences for Pauli limiting effects upon ${\bm B}_{c2}$  for ${\bm B}$ parallel to the low mass direction(s) of  clean, highly anisotropic
superconductors, and the temperature dependence of  Knight shift measurements.  We encourage measurements at higher fields and lower $T$ values to confirm our prediction
that ${\bm B}_{c2,||}(0)$ could greatly exceed the standard Pauli limit in clean monolayer and bilayer superconductors, such as gated and pure transition metal
dichalcogenides \cite{Xi,Lu}.  In monolayer NbSe$_2$, the $g$ factor for ${\bm B}$ parallel to the layers appears to be less than 0.3 \cite{Xi}.  For bilayer and trilayer NbSe$_2$, the analogous $g$-factor is about 0.5-0.7, \cite{Xi} and the actual Pauli limit for twisted bilayer graphene is predicted to already be greatly exceeded by the data, so that $g$ for ${\bm B}$ parallel to the twisted bilayers is most likely on the order of 0.1 or less, since the data are consistent with the top curve in Fig. 5\cite{Cao}.

Recently, it has come to our attention of preprints studying three-layer  stanene and four-layer  PbTe$_2$ \cite{Falson,Liu}.  The resistive transitions are broad, especially for ${\bm B}$ parallel to the films, but the data are consistent with a reduced $g$-factor for that field direction, without having to invoke other mechanisms.  Although the authors interpreted their data as providing evidence for Type-II Ising superconductivity, their data  also support our theory that ultra-thin superconducting layers have a greatly reduced $g$-factor for that field direction.  Phenomenologically, the temperature dependence of $B_{c2,||}(T)$ in monolayer
superconductors should behave as in the Tinkham thin film model \cite{Klemmbook,Tinkham}, but a microscopic theory that does not involve spin-orbit scattering or Ising pairing of
$B_{c2,||}(T)$ in a clean two-dimensional superconductor is sorely needed \cite{KLB,Cao,Xi,Lu,Fatemi,Sajadi,Falson,Liu}.

In addition, the quantum spin Hall Hamiltonian is given for this model by the first of the three parts in the last term in Eq. (28), plus corrections of $O({\rm m}c^2)^{-4}$ given in the Appendix.  In a 2D metal, it is Eq.(\ref{H2DQSH}).  We emphasize that in many papers on the quantum spin Hall effect, the ${\bm A}$ term was omitted, and therefore the full ramifications  of the quantum spin Hall effect have not yet been observed \cite{QiZhang}.  We further note that the coefficient $\mu_B/(2m_{||}c^2)$ is not a free parameter, as $\mu_B=q\hbar/(2{\rm m})$ is the Bohr magneton for a hole and minus the Bohr magneton for an electron, and $m_{||}$ is the particle's planar effective mass that is  measurable for any 2D metal by cyclotron resonance with ${\bm B}$ normal to the conducting plane.

\section{Acknowledgments}
The authors acknowledge helpful discussions with Luca Argenti, Thomas Bullard, Kazuo Kadowaki, and Jingchuan Zhang.
This work was supported by the National Natural Science Foundation of China through Grant no. 11874083. A. Z. was also supported by the China Scholarship Council.  R. A. K. was partially supported by the U. S. Air Force Office of Scientific Research (AFOSR) LRIR \#18RQCOR100, and the AFRL/SFFP Summer Faculty Program provided by AFRL/RQ at WPAFB.

\section{Appendix}
\subsection{Transformations for a constant ${\bm B}$}
The transformations in Eqs. (7, 8, 11-13, 15, 16) apply for a general spatial dependence of ${\bm A}$ and ${\bm B}$.  But for ${\bm B}$ in a constant direction, ${\bm B}=B(\sin\theta\cos\phi,\sin\theta\sin\phi,\cos\theta)$, one can make a further isotropic scale transformation that preserves the scaled Dirac equation.  We set
\begin{eqnarray}
\frac{\partial}{\partial x^{\mu'}}&=&\zeta\frac{\partial}{\partial x^{\mu''}},\nonumber\\
A^{\mu'}&=&\zeta A^{\mu''},\nonumber\\
B^{\mu'}&=&\zeta^2B^{\mu''},\nonumber
\end{eqnarray}
and force $|{\bm B}''|=B$, as sketched in Fig. 4.  This results in $\zeta^{-2}\equiv\alpha(\theta,\phi)$, where
\begin{eqnarray}
\alpha &=&\frac{{\rm m}}{m_g}\sqrt{\frac{m_1}{m_g}\sin^2\theta\cos^2\phi+\frac{m_2}{m_g}\sin^2\theta\sin^2\phi+\frac{m_3}{m_g}\cos^2\theta},
\nonumber
\end{eqnarray}
which differs by ${\rm m}/m_g$ from the scale factor  obtained from the Klemm-Clem transformations \cite{Klemmbook,KlemmClem}, since it preserves the spatial isotropy of the transformed Dirac equation.

\subsection{\label{sec:level2} Proof of covariance for anisotropic Dirac equation}

 To demonstrate the Lorentz invariance of this anisotropic Dirac equation, we employ its contravariant form, multiplied by $\beta/c$ on the left of Eq. (4),
\begin{eqnarray}
\tilde{\gamma}^{0}&=&\beta=\left[\begin{array}{cc}1&0\\0&-1\end{array}\right],\hskip15pt\tilde{\gamma}^{\mu}=\beta\tilde{\alpha}^{\mu}=\left[\begin{array}{cc}0&\tilde{\sigma}^{\mu}\\
-\tilde{\sigma}^{\mu}&0\end{array}\right],\>\>\>\>
\end{eqnarray}
for $\mu=1,2,3$.  We note that $\tilde{\gamma}^0$ is Hermitian, so that $(\tilde{\gamma}^0)^2=1\equiv \tilde{g}^{00}$.  The $\tilde{\gamma}^{\mu}$ for $\mu=1,2,3$ satisfy
the anticommutation
relations
\begin{eqnarray}
\left\{\tilde{\gamma}^{\mu},\tilde{\gamma}^{\nu}\right\}&\equiv&2\tilde{g}^{\mu\nu}\delta^{\mu\nu}=\frac{-2\delta^{\mu\nu}{\rm m}}{m_{\mu}}.
\end{eqnarray}
 These features lead to the metric $\tilde{g}$ given by
\begin{eqnarray}
\tilde{g}&=&\left(\begin{array}{cccc}1&0&0&0\\
0&-{\rm m}/m_1&0&0\\
0&0&-{\rm m}/m_2&0\\
0&0&0&-{\rm m}/m_3\end{array}\right).
\end{eqnarray}
We then may use the Feynman slash notation \cite{BjorkenDrell},
\begin{eqnarray}
\tilde{\slashed{\nabla}}&=&\tilde{\gamma}^{\mu}\frac{\partial}{\partial x^{\mu}}=\frac{\tilde{\gamma}^0}{c}\frac{\partial}{\partial
t}+\tilde{\bm\gamma}\cdot{\bm\nabla},\nonumber\\
\tilde{\slashed{A}}&=&\tilde{\gamma}^{\mu}A_{\mu}=\tilde{\gamma}^0A^0-\tilde{\bm\gamma}\cdot{\bm A},
\end{eqnarray}
to write the anisotropic Dirac equation in covariant form,
\begin{eqnarray}
(i\hbar \tilde{\slashed{\nabla}}-q \tilde{\slashed{A}}-{\rm m}c)\psi&=&0.
\end{eqnarray}

We then employ the Klemm-Clem transformations in the form appropriate for the Dirac equation, as described in detail in the text.
Hence,  the fully general transformed covariant form of the anisotropic Dirac equation for general is
\begin{eqnarray}
(i\hbar \slashed{\nabla}'-q \slashed{A}'-{\rm m}c)\psi&=&0,
\end{eqnarray}
where
\begin{eqnarray}
\slashed{\nabla}'&=&\gamma^{\mu}\frac{\partial}{\partial x^{\mu'}}=\frac{\gamma^0}{c}\frac{\partial}{\partial t}+{\bm\gamma}\cdot{\bm\nabla}',\nonumber\\
\slashed{A}'&=&\gamma^{\mu}A_{\mu}'=\gamma^0A^0-{\bm\gamma}\cdot{\bm A}',\\
\gamma^{0}&=&\beta=\left[\begin{array}{cc}1&0\\0&-1\end{array}\right],\hskip15pt\gamma^{\mu}=\beta\alpha^{\mu}=\left[\begin{array}{cc}0&\sigma^{\mu}\\
-\sigma^{\mu}&0\end{array}\right],\>\>\>\>
\end{eqnarray}
and the transformed metric is identical to that of an isotropic system, with $g^{00}=1$ and $g^{\mu\nu}=-\delta^{\mu\nu}$ for $\mu=1,2,3$.
Thus, Eq. (9) has exactly the same form as does the isotropic covariant form of the Dirac equation, except for the transformed spatial variables.   Hence, the proof of
covariance of the anisotropic Dirac equation under general proper rotations and boosts and under  improper  reflections, charge conjugation, and time reversal
transformations, all follow by inspection from the proofs of the covariance of the isotropic Dirac equation \cite{BjorkenDrell,Jackson}.

\subsection{General proper Lorentz transformations}

For a general proper Lorentz transformation in a relativistic orthorhombic system,
$x'=\tilde{a}x$, where $x'$ and $x$ are column  (Nambu) four-vectors and $\tilde{a}$ is the appropriate proper anisotropic Lorentz transformation, which is to be found
based upon symmetry arguments.  We require the norm with $\tilde{g}$ to be invariant under all possible Lorentz transformations \cite{Jackson}
$(x,\tilde{g}x)=(x',\tilde{g}x')$,
or
$x^{T}\tilde{g}x=(x')^{T}\tilde{g}x'$,
where $x^T$ is the transpose (row) form of the four-vector $x$ and  $\tilde{g}$ is given by Eq. (31).
We then have
$x^T\tilde{g}x=(x')^T\tilde{g}x'=x^T\tilde{a}^T\tilde{g}\tilde{a}x$,
 which implies
$\tilde{g}=\tilde{a}^T\tilde{g}\tilde{a}$.
As for the isotropic case, we assume
$\tilde{a}=e^{\tilde{L}}$,
so that
$\tilde{a}^T=e^{\tilde{L}^T}$, and
$\tilde{a}^{-1}=e^{-\tilde{L}}$.
Then from $\tilde{g}=\tilde{a}^T\tilde{g}\tilde{a}$, we have $\tilde{g}\tilde{a}^{-1}=\tilde{a}^T\tilde{g}$ and hence that
$\tilde{a}^{-1}=\tilde{g}^{-1}\tilde{a}^T\tilde{g}$.  We then may rewrite this as
\begin{eqnarray}
e^{-\tilde{L}}&=&\tilde{g}^{-1}e^{\tilde{L}^T}\tilde{g}=e^{\tilde{g}^{-1}\tilde{L}^T\tilde{g}}.
\end{eqnarray}
 Taking the logarithm of both sides, we obtain
$-\tilde{L}=\tilde{g}^{-1}\tilde{L}^T\tilde{g}$, or that
$-\tilde{g}\tilde{L}=\tilde{L}^T\tilde{g}=(\tilde{g}\tilde{L})^T$,
which requires  $\tilde{g}\tilde{L}$ to be  antisymmetric.  We then write \cite{Jackson}
\begin{eqnarray}
\tilde{L}&=&\left(\begin{array}{cccc}0&\frac{-\zeta_1}{\sqrt{\tilde{m}_1}}&\frac{-\zeta_2}{\sqrt{\tilde{m}_2}}&\frac{-\zeta_3}{\sqrt{\tilde{m}_3}}\\
&&&\\
-\zeta_1\sqrt{\tilde{m}_1}&0&\frac{\omega_3\sqrt{m_1}}{\sqrt{m_2}}&\frac{-\omega_2\sqrt{m_1}}{\sqrt{m_3}}\\
&&&\\
-\zeta_2\sqrt{\tilde{m}_2}&\frac{-\omega_3\sqrt{m_2}}{\sqrt{m_1}}&0&\frac{\omega_1\sqrt{m_2}}{\sqrt{m_3}}\\
&&&\\
-\zeta_3\sqrt{\tilde{m}_3}&\frac{\omega_2\sqrt{m_3}}{\sqrt{m_1}}&\frac{-\omega_1\sqrt{m_3}}{\sqrt{m_2}}&0\end{array}\right),\nonumber\\
\end{eqnarray}
for which $\tilde{g}\tilde{L}$ is easily shown to be antisymmetric.

We then write
\begin{eqnarray}
\tilde{L}&=&-{\bm\omega}\cdot{\tilde{\bm S}}-{\bm\zeta}\cdot{\tilde{\bm K}},
\end{eqnarray}
where
\begin{eqnarray}
\tilde{K}_1&=&\left(\begin{array}{cccc}0&\tilde{m}_1^{-1/2}&0&0\\
\tilde{m}_1^{1/2}&0&0&0\\
0&0&0&0\\
0&0&0&0\end{array}\right),\\
\tilde{K}_2&=&\left(\begin{array}{cccc}0&0&\tilde{m}_2^{-1/2}&0\\
0&0&0&0\\
\tilde{m}_2^{1/2}&0&0&0\\
0&0&0&0\end{array}\right),\\
\tilde{K}_3&=&\left(\begin{array}{cccc}0&0&0&\tilde{m}_3^{-1/2}\\
0&0&0&0\\
0&0&0&0\\
\tilde{m}_3^{1/2}
&0&0&0\end{array}\right),\\
\tilde{S}_1&=&\left(\begin{array}{cccc}0&0&0&0\\
0&0&0&0\\
0&0&0&-\sqrt{\frac{m_2}{m_3}}\\
0&0&\sqrt{\frac{m_3}{m_2}}&0\end{array}\right),\\
\tilde{S}_2&=&\left(\begin{array}{cccc}0&0&0&0\\
0&0&0&\sqrt{\frac{m_1}{m_3}}\\
0&0&0&0\\
0&-\sqrt{\frac{m_3}{m_1}}&0&0\end{array}\right),\\
\tilde{S}_3&=&\left(\begin{array}{cccc}0&0&0&0\\
0&0&-\sqrt{\frac{m_1}{m_2}}&0\\
0&\sqrt{\frac{m_2}{m_1}}&0&0\\
0&0&0&0\end{array}\right).
\end{eqnarray}
It is easy to show that
$\left[\tilde{S}_i,\tilde{S}_j\right]=\epsilon_{ijk}\tilde{S}_k$,
$\left[\tilde{K}_i,\tilde{K}_j\right]=-\epsilon_{ijk}\tilde{S}_k$, and
$\left[\tilde{S}_i,\tilde{K}_j\right]=\epsilon_{ijk}\tilde{K}_k$,
so the anisotropic Lorentz transformation matrix $\tilde{L}$ has SL(2,C) or O(3,1) group symmetry, precisely as for the isotropic case \cite{Jackson}.
We now  show some  examples.  We first define
$\omega=\sqrt{\omega_1^2+\omega_2^2+\omega_3^2}$ and then write
\begin{eqnarray}
A_i&=&\cos\omega+\frac{\omega_i^2}{\omega^2}(1-\cos\omega),\\
B_{ijk}^{\pm}&=&\Bigl(\frac{m_i}{m_j}\Bigr)^{1/2}\Bigl[\frac{\omega_i\omega_j}{\omega^2}(1-\cos\omega)\pm\frac{\omega_k}{\omega}\sin\omega\Bigr].
\end{eqnarray}
Then, for a general rotation,
\begin{eqnarray}
e^{-{\bm\omega}\cdot{\tilde{\bm S}}}&=&
\left(\begin{array}{cccc}1&0&0&0\\
0&A_1&B^{+}_{123}&B^{-}_{132}\\
0&B_{213}^{-}&A_2&B^{+}_{231}\\
0&B^{+}_{312}&B^{-}_{321}&A_3\end{array}\right),
\end{eqnarray}
the determinant of which is 1, as required.

For the general boost case, we first set ${\bm \zeta}=\hat{\overline{\bm \beta}}\tanh^{-1}\overline{\beta}$, where $\overline{\bm\beta}={\bm v}/c$, ${\bm v}$ is the
electron's velocity, and define
$\zeta=\sqrt{\zeta_1^2+\zeta_2^2+\zeta_3^2}$,
$\cosh\zeta=\gamma=\frac{1}{\sqrt{1-\overline{\beta}^2}}$,
$\sinh\zeta=\gamma\overline{\beta}$, and
$\overline{\beta}=\sqrt{\overline{\beta}_1^2+\overline{\beta}_2^2+\overline{\beta}_3^2}$, as for the isotropic case \cite{Jackson}.  Then we define
\begin{eqnarray}
C_{i}^{\pm}&=&-\gamma\overline{\beta}_i\tilde{m}_i^{\pm 1/2},\\
D_{i}&=&1+\frac{(\gamma-1)\overline{\beta}_i^2}{\overline{\beta}^2},\\
E_{ij}&=&(\gamma-1)\frac{\overline{\beta}_i\overline{\beta}_j}{\overline{\beta}^2}\Bigl(\frac{m_i}{m_j}\Bigr)^{1/2}.
\end{eqnarray}
Then for the general boost case, we have \cite{Jackson}
\begin{eqnarray}
e^{-{\bm\zeta}\cdot\tilde{\bm K}}
&=&\left(\begin{array}{cccc}\gamma&C_1^{-}&C_2^{-}&C_3^{-}\\
C_1^{+}&D_1&E_{12}&E_{13}\\
C_2^{+}&E_{21}&D_2&E_{23}\\
C_3^{+}&E_{31}&E_{32}&D_3\end{array}\right),
\end{eqnarray}
the determinant of which is  also 1,
as required.

Hence  Eq. (4) is invariant under the most general proper Lorentz transformation.  As argued in the following, it is also invariant under all of the
relevant improper Lorentz transformations:  reflections or parity, charge conjugation, and time reversal \cite{BjorkenDrell}.

\subsection{\label{sec:level2} Covariance under reflections  by an arbitrary plane}
A general reflection about a plane normal to the $z$ axis may be written as a rank-4 matrix as
\begin{eqnarray}
R&=&\left(\begin{array}{cccc} 1&0&0&0\\
0&\cos\alpha&-\sin\alpha&0\\
0&\sin\alpha&\cos\alpha&0\\
0&0&0&-1\end{array}\right),
\end{eqnarray}
the determinant of which is $|R|=-1$ \cite{Boas}.  It is elementary to change this to a general reflection about either the $x$ or the $y$ axis.
Then, we may combine a general rotation given by the matrix $e^{-{\bm\omega}\cdot\tilde{S}}$ with such a reflection as that described by $R$, either as $Re^{-{\bm\omega}\cdot\tilde{S}}$ or as $e^{-{\bm\omega}\cdot\tilde{S}}R$.  Although these two matrices do not in general commute, their combined determinant $|Re^{-{\bm\omega}\cdot\tilde{S}}|=|e^{-{\bm\omega}\cdot\tilde{S}}R|=|R||e^{-{\bm\omega}\cdot\tilde{S}}|=-1$ \cite{Nering}.  Similarly, the theorem also implies that the determinant of a combined boost and reflection $|Re^{-{\bm\zeta}\cdot\tilde{K}}|=|e^{-{\bm\zeta}\cdot\tilde{K}}R|=-1$.  Hence, this is precisely the same as for the isotropic Dirac equation in 3D.

\subsection{\label{sec:level2} Covariance under CPT transformations}
As for the isotropic case, reflections require ${\bm x}'=-{\bm x}$
and $t'=t$, so that $b$ is a diagonal rank-4 matrix with $b^{00}=1$ and $b^{\mu\mu}=-1$ for $\mu=1,2,3$, which is identical to $\tilde{g}$ in the isotropic limit
\cite{BjorkenDrell}. Reflections can then be represented by a unitary  matrix $P$ satisfying
\begin{eqnarray}
P^{-1}\tilde{\gamma}^{\mu}P&=&b^{\mu\mu}\tilde{\gamma}^{\mu},
\end{eqnarray}
which is satisfied for
\begin{eqnarray}
P&=&e^{i\phi}\tilde{\gamma}^0,
\end{eqnarray}
where the phase factor $\phi=n\pi/2$ for integer $n$, so that four reflections leaves $\psi$ invariant, as for a rotation through $4\pi$ about the quantization axis of a
spin 1/2 spinor.  We also have that $P\Phi({\bm x},t)=\Phi'({\bm x}',t)=\Phi({\bm x},t)$ is even under parity, and $P{\bm A}({\bm x},t)={\bm A}'({\bm x}'.t)=-{\bm A}({\bm
x},t)$ is odd under parity.

As for charge conjugation, the hole wave function $\psi_c$ in an anisotropic conduction band  satisfies
\begin{eqnarray}
(i\hbar\tilde{\slashed{\nabla}}-e\tilde{\slashed{A}}-mc)\psi_c&=&0.\label{positronDirac}
\end{eqnarray}
This is accomplished by taking the complex conjugate:
\begin{eqnarray}
\psi_c&=&C\overline{\psi}^T,
\end{eqnarray}
where $C=i\gamma^2\gamma^0$. As for the isotropic Dirac equation, charge conjugation  is also given by  $C=i\gamma^2\gamma^0$, so that
\begin{eqnarray}
C^{-1}\tilde{\gamma}^{\mu}C&=&-\tilde{\gamma}^{\mu T}.
\end{eqnarray}

Under time reversal, we have the properties that if $t'=-t$,
$\psi'(t')=T\psi^{*}(t)$
where
\begin{eqnarray}
T&=&i\gamma^1\gamma^3.
\end{eqnarray}
Hence, Eq. (4) is invariant under  charge conjugation, parity, and time-reversal (CPT) transformations.

\subsection{Foldy-Wouthuysen transformations to fourth order}
For an orthorhombically-anisotropic conduction band, the Dirac Hamiltonian ${\tilde H}$ is
written in Eq. (4) of the text, and the Schr{\"o}dinger equation is $\tilde{H}\psi=i\hbar\frac{\partial\psi}{\partial t}$.  As shown previously \cite{FW,BjorkenDrell}, the correct expansion about the non-relativistic limit removes terms linear in $\tilde{\cal O}$ by the transformation
\begin{eqnarray}
\psi'&=&e^{i\tilde{S}}\psi,
\end{eqnarray}
leading to the transformed Schr{\"o}dinger equation
\begin{eqnarray}
\tilde{H}'\psi'&=&i\hbar\frac{\partial \psi'}{\partial t},
\end{eqnarray}
where

\begin{eqnarray}
\tilde{H}'&=&e^{i\tilde{S}}\tilde{H}e^{-i\tilde{S}}-i\hbar e^{i\tilde{S}}\Bigl(\frac{\partial}{\partial t}e^{-i\tilde{S}}\Bigr).
\end{eqnarray}
To  order $({\rm m}c^2)^{-4}$, we require

\begin{align*}
\tilde{H}^{\prime}&=\tilde{H}+i[\tilde{S},\tilde{H}]+\frac{i^2}{2!}\big[\tilde{S},[\tilde{S},\tilde{H}]\big]+\frac{i^3}{3!}\bigl[\tilde{S},[\tilde{S},[\tilde{S},\tilde{H}]]\bigr]\\
& +\frac{i^4}{4!}\bigl[\tilde{S},\bigl[\tilde{S},[\tilde{S},[\tilde{S},\tilde{H}]]\bigr]\bigr]
+\frac{i^5}{5!}\bigl[\tilde{S},\bigl[\tilde{S},\bigl[\tilde{S},[\tilde{S},[\tilde{S},\tilde{H}]]\bigr]\bigr]\bigr]  \\
&+\frac{i^6}{6!}\bigl[\tilde{S},\bigl[\tilde{S},\bigl[\tilde{S},\bigl[\tilde{S},[\tilde{S},[\tilde{S},\tilde{H}]]\bigr]\bigr]\bigr]\bigr]-\hbar\frac{\partial{\tilde{S}}}{\partial t}
-\frac{i\hbar}{2!}[\tilde{S},\frac{\partial{\tilde{S}}}{\partial t}]  \\
&-\frac{i^2 \hbar}{3!}\bigl[\tilde{S},[\tilde{S},\frac{\partial{\tilde{S}}}{\partial t}]\bigr]-\frac{i^3 \hbar}{4!}\bigl[\tilde{S},[\tilde{S},[\tilde{S},\frac{\partial{\tilde{S}}}{\partial t}]]\bigr]\\
&-\frac{i^4 \hbar}{5!}\bigl[\tilde{S},\bigl[\tilde{S},[\tilde{S},[\tilde{S},\frac{\partial{\tilde{S}}}{\partial t}]]\bigr]\bigr],
\end{align*}
where $\tilde{S}=-\frac{i\beta \tilde{\cal O}}{2{\rm m}c^2}$.

\subsection{\label{sec:level2}Terms to fourth order in $1/({\rm m}c^2)$}
The anisotropic 3D Hamiltonian to order $({\rm m}c^2)^{-4}$ in untransformed space is

\begin{eqnarray}
\tilde{H}_{3D}^{NR}&=& \beta\Biggl({\rm m}c^2+\sum_{\mu=1}^3\bigl(\frac{\Pi_{\mu}^2}{2m_{\mu}}
-\frac{\mu_B{\rm m}\sqrt{m_{\mu}}\sigma_{\mu}B_{\mu}}{(m_g)^{3/2}}\bigr)  \nonumber\\
&&-\frac{1}{2{\rm m}c^2}\bigl(\sum_{\mu=1}^3\frac{\Pi_{\mu}^2}{2m_{\mu}}
-\frac{\mu_B{\rm m}\sqrt{m_{\mu}}\sigma_{\mu}B_{\mu}}{(m_g)^{3/2}}\bigr)^2 \nonumber\\
&&+\frac{\mu_B^2}{2c^4}\sum_{\mu=1}^3\frac{E_{\mu}^2}{m_{\mu}}-\frac{i\hbar^3q^2}{32{\rm m}^4c^6}\sum_{\mu=1}^3\frac{{\rm m}}{m_{\mu}}\frac{\partial}{\partial t}E_{\mu}^2\Biggr)\nonumber\\
&&-\frac{\mu_B}{4c^2}\sum_{\mu=1}^3\frac{\hbar}{m_{\mu}}\frac{\partial E_{\mu}}{\partial x_{\mu}}+q\Phi({\bm r},t)\nonumber\\
&&-\frac{\mu_B}{4c^2(m_g)^{3/2}}   \sum_{\mu,\nu,\lambda=1}^3\Bigl((2E_{\mu}\Pi_{\nu}-i\hbar\frac{\partial E_{\mu}}{\partial x_{\nu}})  \nonumber\\
&& \times \epsilon_{\mu\nu\lambda}\sqrt{m_{\lambda}}\sigma_{\lambda}\Bigr)
+\frac{\tilde{Z}_{3D}}{384{\rm m}^4 c^8}+O({\rm m}c^2)^{-5},\label{HNR}
\end{eqnarray}
\vskip5pt
where $\tilde{Z}_{3D}= [\tilde{{\cal O}},[\tilde{{\cal O}},[\tilde{{\cal O}}, \tilde{\Gamma}]]]$ is given by
\begin{eqnarray}
\tilde{Z}_{3D}&=&i\hbar c^4q\sum_{\mu\nu}\frac{{\rm m}^2}{m_{\mu}m_{\nu}}[\Pi_{\nu},M_{\nu\mu}]\nonumber\\
& &-\frac{\hbar c^4q{\rm m}^2}{m_g^{3/2}}\sum_{\mu\nu\lambda\delta}\sigma_{\delta}\Biggl(\frac{m_{\delta}^{1/2}\epsilon_{\lambda\nu\delta}}{m_{\mu}}
\{\Pi_{\lambda},M_{\nu\mu}\}\nonumber\\
& &+\frac{\epsilon_{\mu\nu\lambda}}{m_{\delta}^{1/2}}\Bigl[\Pi_{\delta},[\Pi_{\lambda},\{\Pi_{\mu},E_{\nu}\}]\Bigr]\Biggr),\\
M_{\nu\mu}&=&\Bigl[\Pi_{\nu},[\Pi_{\mu},E_{\mu}]\Bigr]-\Bigl\{\Pi_{\mu},\{\Pi_{\nu},E_{\mu}\}-\{\Pi_{\mu},E_{\nu}\}\Bigr\}\nonumber\\
&=&\Bigl\{\Pi_{\mu},\{\Pi_{\mu},E_{\nu}\}\Bigr\}-2\Bigl(\Pi_{\mu}E_{\mu}\Pi_{\nu}+\Pi_{\nu}E_{\mu}\Pi_{\mu}\Bigr),\nonumber\\
\end{eqnarray}
where we made use of the fact that
$[E_{\mu},[\Pi_{\mu},\Pi_{\nu}]]=0$,
since for $\mu=\nu$, $[\Pi_{\mu},\Pi_{\mu}]=0$, and for $\mu\ne\nu$, $[E_{\mu},[\Pi_{\mu},\Pi_{\nu}]]=i\sum_{\gamma}\hbar q\epsilon_{\mu\nu\gamma}[E_{\mu},B_{\gamma}]=0$.

In order to obtain a useful expression for $\tilde{Z}_{3D}$, it is helpful to first perform the Klemm-Clem transformations in Eqs. (7),(8), (11-13), (15), and (16) on $\tilde{H}_{3D}$ to $O({\rm m}c^2)^{-4}$.  We obtain

\begin{eqnarray}
H'_{3D}&=&\beta\Biggl({\rm m}c^2+\frac{{\bm \Pi'}^2}{2{\rm m}}-\mu_{B}{\bm \sigma}\cdot{\bm B}'+\frac{\mu_B^2}{2{\rm m}c^4}{{\bm E}'}^2 \nonumber \\
&&-\frac{1}{2{\rm m}c^2}\Bigl(\frac{{\bm \Pi'}^2}{2{\rm m}}-\mu_{B}{\bm \sigma}\cdot{\bm B}'\Bigr)^2-\frac{i\hbar\mu_B^2}{8{\rm m}^2c^6}\frac{\partial}{\partial t}({\bm E}')^2\Biggr)\nonumber\\
&&-\frac{\mu_{B}}{4{\rm m}c^2}\Bigl(\hbar{\bm \nabla}'\cdot{\bm E}'+(2{\bm E}'\times{\bm\Pi}'+i\hbar{\bm\nabla}'\times{\bm E}')\cdot{\bm\sigma}\Bigr)\nonumber\\
& &+\frac{\mu_B}{192{\rm m}^3c^4}\sum_{i=1}^3X'_{3D,i}
+q\Phi({\bm r}',t)+O({\rm m}c^2)^{-5},\nonumber
\end{eqnarray}
where
\begin{eqnarray}
X'_{3D,1}&=&-({\bm\Pi}'\times{\bm M}'-{\bm M}'\times{\bm\Pi}')\cdot{\bm\sigma},\\
X'_{3D,2}&=&-\Bigl[{\bm\sigma}\cdot{\bm\Pi}',({\bm\Pi}'\times{\bm\Pi}')\cdot{\bm E}'\nonumber\\
&&-2({\bm\Pi}'\times{\bm E}')\cdot{\bm\Pi}'
+({\bm E}'\times{\bm\Pi}')\cdot{\bm\Pi}'\Bigr],\nonumber\\
X'_{3D,3}&=&i({\bm\Pi}'\cdot{\bm M}'-{\bm M}'\cdot{\bm \Pi}'),
\end{eqnarray}
where ${\bm M}'={\bm M}_1'+{\bm M}_2'$, and
\begin{eqnarray}
{\bm M}_1'&=&-2({\bm \Pi}'\cdot{\bm E}'){\bm\Pi}'-2{\bm\Pi}'({\bm E}'\cdot{\bm\Pi}'),\nonumber\\
{\bm M}_2'&=&2{\bm \Pi}'^2{\bm E}'+2{\bm E}'{\bm\Pi}'^2+\hbar^2{\bm\nabla}'^2{\bm E}'.\label{Mprime}
\end{eqnarray}
After some algebra, the fourth order terms $X_i'$  are then found to be $X'_{3D}=\sum\limits_{i}X'_{3D,{i}}$ and
\begin{eqnarray}
X'_{3D,(1)}&=&-8i\hbar q[({\bm E}'\cdot{\bm B}')({\bm\sigma}\cdot{\bm\Pi}')-({\bm\sigma}\cdot{\bm B}')({\bm E}'\cdot{\bm\Pi}')],\nonumber\\
X'_{3D,(2)}&=&-2\hbar^2q[({\bm E}'\cdot{\bm\nabla}')({\bm\sigma}\cdot{\bm B}')],\nonumber\\
X'_{3D,(3)}&=&4\hbar^2q[{\bm E}'\times({\bm\nabla}'\times{\bm B}')]\cdot{\bm\sigma},\nonumber\\
X'_{3D,(4)}&=&-4\hbar^2q\Bigl({\bm B}'\cdot[({\bm \sigma}\cdot{\bm\nabla}'){\bm E}']+({\bm B}'\cdot{\bm\nabla}')({\bm E}'\cdot{\bm\sigma})\Bigr),\nonumber\\
X'_{3D,(5)}&=&+8\hbar^2q({\bm\sigma}\cdot{\bm B}')({\bm\nabla}'\cdot{\bm E}'),\nonumber\\
X'_{3D,(6)}&=&-4\hbar^2q({\bm\sigma}\cdot{\bm\nabla}')({\bm B}'\cdot{\bm E}'),\nonumber\\
X'_{3D,(7)}&=&4i\hbar[({\bm\nabla}'\times{\bm E}')\cdot{\bm \sigma}]{\bm\Pi}'^2,\nonumber\\
X'_{3D,(8)}&=&+4\hbar^2\Bigl({\bm\nabla}'[({\bm\nabla}'\times{\bm E}')\cdot{\bm\sigma}]\Bigr)\cdot{\bm\Pi}',\nonumber\\
X'_{3D,(9)}&=&-2\hbar^2q({\bm\sigma}\times{\bm B}')\cdot({\bm\nabla}'\times{\bm E}'),\nonumber\\
X'_{3D,(10)}&=&8[({\bm E}'\times{\bm\Pi}')\cdot{\bm\sigma}]{\bm\Pi}'^2,\nonumber\\
X'_{3D,(11)}&=&-2\hbar^2[({\bm\nabla}'^2{\bm E}')\times{\bm\Pi}']\cdot{\bm\sigma},\nonumber\\
X'_{3D,(12)}&=&-i\hbar^3[{\bm\nabla}'^2({\bm\nabla}'\times{\bm E}')]\cdot{\bm\sigma},\nonumber\\
X'_{3D,(13)}&=&-2\hbar^2[{\bm\nabla}'({\bm\nabla}'\cdot{\bm E}')\times{\bm\Pi}']\cdot{\bm\sigma},\nonumber\\
X'_{3D,(14)}&=&-2\hbar^2\Bigl[({\bm\sigma}\cdot{\bm\nabla}')({\bm\nabla}'\times{\bm E}')\Bigr]\cdot{\bm\Pi}',\nonumber\\
X'_{3D,(15)}&=&-8i\hbar\sum_{\mu}{\bm\sigma}\cdot\Bigl(\frac{\partial{\bm E}'}{\partial x_{\mu}'}\times{\bm\Pi}'\Bigr){\bm\Pi}_{\mu}',\nonumber\\
\end{eqnarray}
and since $X_{3D,3}'$ does not depend upon the spin, it does not contribute to the Zeeman, spin-orbit, and the quantum spin Hall interactions.

For an electron or hole in a 2D metal with $m_1=m_2=m_{||}$ and $m_3\rightarrow\infty$, the planar-isotropic 2D Hamiltonian to order $({\rm m}c^2)^{-4}$ in untransformed space is
\begin{eqnarray}
H_{2D}&=&\beta\Biggl({\rm m}c^2+\frac{{\bm \Pi}^{'2}_{||}}{2{\rm m}}-\mu_{B}\sigma_{\perp}B'_{\perp}+\frac{\mu_{B}^2}{2{\rm m}c^4}{\bm E}^{'2}_{||} \nonumber \\
&&-\frac{1}{2{\rm m}c^2}\Bigl(\frac{{\bm \Pi}^{'2}_{||}}{2{\rm m}}-\mu_{B}\sigma_{\perp}B'_{\perp}\Bigr)^2-\frac{i\hbar\mu_B^2}{8{\rm m}^2c^6}\frac{\partial}{\partial t}\Bigl({\bm E}^{'2}_{||}\Bigr)\Biggr)\nonumber\\
&&-\frac{\mu_{B}}{4{\rm m}c^2}\Bigl(\hbar{\bm \nabla}'_{||}\cdot{\bm E}'_{||}+\Bigl(2{\bm E}'\times{\bm\Pi}'+i\hbar{\bm\nabla}'\times{\bm
E}'\Bigr)_{\perp}\sigma_{\perp}\Bigr)\nonumber\\
& &+\frac{\mu_BX'_{2D}}{192{\rm m}^3c^4}
+q\Phi({\bm r}_{||},t)+O({\rm m}c^2)^{-5}, \nonumber\\
\end{eqnarray}
where ${\bm M}$ is the unprimed version of ${\bm M}'$  in Eq. (\ref{Mprime}),
 the subscripts $||$ and $\perp$ respectively denote the  components parallel and perpendicular to the film,
$\mu_{B||}=q\hbar/(2m_{||})$ is the 2D effective Bohr magneton for a hole (or minus that for an electron), and

\begin{eqnarray}
X'_{2D}&=&-({\bm\Pi}'\times{\bm M}'-{\bm M}'\times{\bm\Pi}')_{\perp}{\bm\sigma}_{\perp}
   +i({\bm \Pi}'\cdot{\bm M}'-{\bm M}'\cdot{\bm\Pi}')_{||}\nonumber\\
&=&\sum_{i=1}^9X'_{2D,(i)}.
\end{eqnarray}
For $m_1=m_2=m_{||}$, $X_{2D}=\sum_{i=1}^9X_{2D,(i)}$ where
\begin{eqnarray}
X_{2D,(1)}&=&8i\hbar q{\bm\sigma}_{\perp}{\bm B}_{\perp}({\bm E}\cdot{\bm\Pi})_{\parallel},\nonumber\\
X_{2D,(2)}&=&4\hbar^2q({\bm E}\cdot{\bm\nabla})_{\parallel}({\bm\sigma}_{\perp}{\bm B}_{\perp}),\nonumber\\
X_{2D,(3)}&=&4\hbar^2q({\bm\sigma}_{\perp}{\bm B}_{\perp})({\bm\nabla}\cdot{\bm E})_{\parallel},\nonumber\\
X_{2D,(4)}&=&4i\hbar({\bm\nabla}\times{\bm E})_{\perp}{\bm \sigma}_{\perp}{\bm\Pi}_{\parallel}{^2},\nonumber\\
X_{2D,(5)}&=&4\hbar^2{\bm\sigma}_{\perp}\Bigl({\bm\nabla}_{\parallel}[({\bm\nabla}\times{\bm E})]_{\perp}\Bigr)\cdot{\bm\Pi}_{\parallel},\nonumber\\
X_{2D,(6)}&=&8({\bm E}\times{\bm\Pi})_{\perp}{\bm\sigma}_{\perp}{\bm\Pi}_{\parallel}{^2},\nonumber\\
X_{2D,(7)}&=&-2\hbar^2[({\bm\nabla}^2{\bm E})\times{\bm\Pi}]_{\perp}{\bm\sigma}_{\perp},\nonumber\\
X_{2D,(8)}&=&-i\hbar^3[{\bm\nabla}{^2}_{\parallel}({\bm\nabla}\times{\bm E})_{\perp}]{\bm\sigma}_{\perp},\nonumber\\
X_{2D,(9)}&=&-8i\hbar\sigma_{\perp}\sum_{\mu=1}^2\Bigl(\frac{\partial{\bm E}}{\partial x_{\mu}}\times{\bm\Pi}\Bigr)_{\perp}\Pi_{\mu},
\end{eqnarray}
and the remaining terms arising from $i({\bm\Pi}\cdot{\bm M}-{\bm M}\cdot{\bm\Pi})$ do not depend upon the spin.

\begin{thebibliography}{99}
\bibitem{Feynman} R. P. Feynman, R. B. Leighton, and M. Sands, {\it The Feynman Lectures on Physics, Vol. II} (Addison-Wesley, Reading, MA, 1964).
\bibitem{Griffiths} D. J. Griffiths and D. J. Schroeter, {\it Introduction to Quantum Mechanics}, (3$^{rd}$ Ed., Cambridge University Press, Cambridge, UK, 2018).
\bibitem{Dirac} P. A. M. Dirac, {\it The Quantum Theory of the Electron}, Proc. Roy. Soc. (London), {\bf A117}, 610 (1928); {\it ibid.} {\bf A118}, 351 (1928).
\bibitem{Cohen} M. L. Cohen and T. K. Bergstresser, {\it Band Structures and Pseudopotential Form Factors for Fourteen Semiconductors of the Diamond and Zinc-blende
    Structures}, Phys. Rev. {\bf 141}, 789 (1966).
\bibitem{CohenBlount} M. H. Cohen and E. I. Blount, {\it The g-factor and de Haas-van Alphen Effect of Electrons in Bismuth}, Phil. Mag. {\bf 5}, 115 (1960).
\bibitem{Mahan} G. D. Mahan, {\it Condensed Matter in a Nutshell}, (Princeton University Press, Princeton, NJ, 2011).
\bibitem{Friedmann} T. A. Friedmann, M. W. Rabin, J. Giapintzakis, J. P. Rice, and D. M. Ginsberg, {\it Direct Measurement of the Anisotropy of the Resistivity in the a-b
    Plane of Twin-Free, Single-Crystal, Superconducting \rm{YBa$_{2}$Cu$_{3}$O$_{7-\delta}$}}, Phys. Rev. B {\bf 42}, 6217 (1990).
\bibitem{Klemmbook} R. A. Klemm, {\it Layered Superconductors, Volume 1} (Oxford University Press, Oxford, UK and New York, NY, 2012).
\bibitem{Barrett} S. E. Barrett, D. J. Durand, C. H. Pennington, C. P. Slichter, T. A. Friedmann, J. P. Rice, and D. M. Ginsberg,  \rm{${}^{63}$Cu {\it Knight Shifts in
    the
    Superconducting State of \rm{YBa$_{2}$Cu$_{3}$O$_{7-\delta}$ (T$_{c}$=90 K)}}}, Phys. Rev. B {\bf 41}, 6283 (1990).
\bibitem{Yosida} Y. Yosida, {\it Paramagnetic susceptibility in superconductors}, Phys. Rev. {\bf 110}, 769 (1958).
\bibitem{Ishida} K. Ishida, H. Mukuda, Y. Kitaoka, K. Asayaa, Z. Q. Mao, Y. Mori, Y. Maeno, {\it  Spin-Triplet Superconductivity in {\rm{Sr$_2$RuO$_4$}} Identified by
    ${}^{17}$O
    Knight Shift}, Nature {\bf 396}, 65 (1998).
\bibitem{Suderow} H. Suderow, V. Crespo, I. Guillamon, S. Vieira, F. Servant, P. Lejay, J. P. Brison, and J. Flouquet, {\it A Nodeless Superconducting Gap in
    {\rm{Sr$_{2}$RuO$_{4}$}} from Tunneling Spectroscopy}, New J. Phys. {\bf 11}, 93004 (2009).
\bibitem{Pustogow} A. Pustogow, Yongkang Luo, A. Chronister, Y.-S. Su, D. A. Sokolov, F. Jerzembeck, A. P. Mackenzie, C. W. Hicks, N. Kikugawa, S. Raghu, E. D. Bauer, S. E.
    Brown, {\it Constraints on the Superconducting Order Parameter in
     {\rm{Sr${_2}$RuO${_4}$}} from Oxygen-17 Nuclear Magnetic Resonance }, Nature {\bf 574}, 72 (2019).
\bibitem{Ishida2} K. Ishida, M. Manago, and Y. Maeno, \rm{$^{17}$O} {\it Knight shift in the Superconducting State and the Heat-up Effect by NMR Pulses on} Sr$_2$RuO$_4$,
    ArXiv: 1907.12236v2 (2019).
\bibitem{Chaikin} I. J. Lee, S. E. Brown, W. G. Clark, M. J. Strouse, M. J. Naughton, W. Kang, and P. M. Chaikin, {\it Triplet Superconductivity in an Organic
    Superconductor Probed by NMR Knight Shift}, Phys. Rev. Lett. {\bf 88}, 17004 (2001).
\bibitem{Cao} Y. Cao, V. Fatemi, S. Fang, K. Watanabe, T. Taniguchi, E. Kaxiras, P. Jarillo-Herrero, {\it Unconventional Superconductivity in Magic-Angle Graphene
    Superlattices},  Nature {\bf 536}, 43-50 (2018).
\bibitem{KlemmClem} R. A. Klemm and J. R. Clem, {\it Lower Critical Field of an Anisotropic Type-\uppercase\expandafter{\romannumeral2} Superconductor}, Phys. Rev. B {\bf 21}, 1868  (1980).
\bibitem{FW} L. L. Foldy and S. A. Wouthuysen, {\it On the Dirac Theory of Spin {\rm{1/2}} Particles and Its Non-relativistic Limit}, Phys. Rev. {\bf 78}, 29-36
    (1950).
\bibitem{Xi} X. Xi, Z. Wang, W. Zhao, J.-H. Park, K. T. Law, H. Berger, L. Forró, J. Shan, K. F. Mak, {\it Ising Pairing in Superconducting {\rm{NbSe$_{2}$}} Atomic
    Layers},
     Nat. Phys. {\bf 12}, 139-143 (2016).
\bibitem{Lu} J. M. Lu, O. Zheliuk, I. Leermakers, N. F. Q. Yuan, U. Zeitler, K. T. Law, J. T. Ye, {\it Evidence for Two-Dimensional Ising Superconductivity in Gated
    {\rm{MoS$_{2}$}} },  Science {\bf 350}, 1353-1357 (2015).
\bibitem{Fatemi} V. Fatemi, S. Wu, Y. Cao, L. Bretheau, Q. D. Gibson, K. Watanabe, T. Taniguchi, R. J. Cava, P. Jarillo-Herrero, {\it Electrically Tunable Low-Density
    Superconductivity in a Monolayer Topological Insulator},  Science {\bf 362}, 926-929 (2018).
\bibitem{Sajadi} E. Sajadi, T. Palomaki, Z. Fei, W. Zhao, P. Bement, C. Olsen, S. Luescher, X. Xu, J. A. Folk, D. H. Cobden, {\it Gate-induced Superconductivity in a
    Monolayer Topological Insulator}, Science {\bf 362}, 922-925 (2018).
\bibitem{Agosta} C. C. Agosta, N. A. Fortune, S. T. Hannahs, S. Gu, L. Liang, J.-H. Park, J. A. Schleuter, {\it Calorimetric Measurements of Magnetic-Field-Induced
    Inhomogeneous Superconductivity above the Paramagnetic Limit},  Phys. Rev. Lett. {\bf 118}, 267001 (2017).
\bibitem{Matsuda} Y. Matsuda, H. Shimahara, {\it Fulde–-Ferrell-–Larkin-–Ovchinnikov State in Heavy Fermion Superconductors}, J. Phys. Soc. Jpn. {\bf 76}, 051005
    (2007).
\bibitem{KLB} R. A. Klemm, A. Luther, M. R. Beasley, {\it Theory of the Upper Critical Field in Layered Superconductors},  Phys. Rev. B {\bf 12}, 877-891 (1975).
\bibitem{FF} P. Fulde, R. A. Farrell, {\it Superconductivity in a Strong Spin-Exchange Field},  Phys. Rev. {\bf 135}, A550-A562 (1964).
\bibitem{LO} A. I. Larkin, Y. N. Ovchinnikov, {\it Nonuniform State of Superconductors},  Zh. Eksp. Teor. Fiz. {\bf 47}, 1136 (1964).
\bibitem{HallKlemm} B. E. Hall and R. A. Klemm, {\it Microscopic Model of the Knight Shift in Anisotropic and Correlated Metals}, J. Phys.: Condens. Matter {\bf 28},
    03LT01 (2016).
\bibitem{magnetochemistry} R. A. Klemm, {\it  Towards a Microscopic Theory of the Knight Shift in an Anisotropic, Multiband Type-II Superconductor}, Magnetochemistry {\bf
    14}, 4, (2018).
\bibitem{Zeinab} Z. El-Moussawi, A. Nourdine, H. Medlej, T. Hamieh, P. Chenevier, and L. Flandin, {\it Fine Tuning of Optoelectronic Properties of Single-Walled Carbon Nanotubes from Conductors to Semiconductors}, Carbon {\bf 153}, 337 (2019).
\bibitem{Dresselhaus} R. Saito, G. Dresselhaus, and M. S. Dresselhaus, {\it Physical Properties of Carbon Nanotubes} (World Scientific, 1998).
\bibitem{Keller} H. J. Keller, ed., {\it Chemistry and Physics of One-Dimensional Metals}, (NATO Advanced Study Institutes Series B: Physics, vol. 25, Plenum, New York, NY, 1976).
\bibitem{1Dconductors}  J. Ehlers, K. Hepp, and H. A. Weidenm{\"u}ller, Eds., {\it One-Dimensional Conductors}, (Lecture Notes in Physics, GPS Summer School Proceedings, Springer, Berlin, 1975).
\bibitem{Hokkaido} Y. Abe, M. Ido, K. Imai, T. Haga, J. Nakahara, T. Sambongi, H. Takayama, S. Tanaka, and K. Yamaya, Eds. {\it Nonlinear Transport and Related Phenomena in Inorganic Quasi One Dimensional Conductors}, (Proceedings of the International Symposium, Hokkaido University, Sapporo, Japan, 1983)
\bibitem{He} S. He, J. He, W. Zhang, L. Zhao, D. Liu, X. Liu, D. Mou, Y.-B. Ou, Q.-Y. Wang, Z. Li, L. Wang, Y. Peng, Y. Liu, C. Chen, L. Yu, G. Liu, X. Dong, J. Zhang, C. Chen, Z. Xu, X. Chen, X. Ma, Q. Xue, and X. J. Zhou, {\it Phase Diagram and Electronic Indication of High-Temperature Superconductivity at 65 K in Single-Layer FeSe Films}, Nature Mat. {\bf 12}, 605 (2013).
\bibitem{KlemmPristine} R. A. Klemm, {\it Pristine and Intercalated Transition Metal Dichalcogenide Superconductors}, Physica C, {\bf 514}, 84 (2015).
\bibitem{Wang} C. Wang, B. Lian, X. Guo, J. Mao, Z. Zhang, D. Zhang, B.-L. Gu, Y. Xu, and W. Duan, {\it Type-II Ising Superconductivity in Two-dimensional Materials with Spin-orbit Coupling}, Phys. Rev. Lett. {bf 123}, 126402 (2019).
\bibitem{BjorkenDrell} J. D. Bjorken and S. D. Drell, {\it Relativistic Quantum Mechanics} (McGraw-Hill, New York, 1964).
\bibitem{Nering} E. D. Nering, {\it Linear Algebra and Matrix Theory}, (Wiley, New York, NY, 1963).
\bibitem{Zhao} A. Zhao, J. Zhang, Q. Gu, and R. A. Klemm, {\it A Relativistic Electron in an Anisotropic Conduction Band}, ArXiv: 1905.03127.
\bibitem{QiZhang} X.-L. Qi and S.-C. Zhang, {\it Topological Insulators and Superconductors}, Rev. Mod. Phys. {\bf 83}, 1057 (2011).
\bibitem{Tinkham} M. Tinkham, {\it Effect of Fluxoid Quantization on Transitions of Superconducting Films}, Phys. Rev. {\bf 129}, 2413 (1963).
\bibitem{Falson} J. Falson, Y. Xu, M. Liao, Y. Zang, K. Zhu, C. Wang, Z. Zhang, H. Liu, W. Duan,  K. He, H. Liu, J. H. Smet, D. Zhang, and Q.-K. Xue, {\it Type-II Ising Pairing in Few-Layer Stanene}, arXiv:1903.07627.
\bibitem{Liu} Y. Liu, Y. Xu, J. Sun, C. Liu, Y. Liu, C. Wang, Z. Zhang, K. Gu, Y. Tang, C. Dang, H. Liu, H. Yao, X. Lin, L. Wang, Q.-K. Xue, and J. Wang, {\it Quantum Metal State and Quantum Phase Transitions in Type-II Ising Superconducting Films}, arXiv: 1904.12719.
\bibitem{Jackson} J. D. Jackson, {\it Classical Electrodynamics}, Third Ed. (Wiley, Hoboken, NJ, 1999), Chapter 11.
\bibitem{Boas} M. L. Boas, {\it Mathematical Methods in the Physical Sciences, 3rd Ed.} (Wiley, Danvers, MA and Hoboken, NJ, 2006).

\end{thebibliography}
\end{document}